\documentclass[useAMS]{mn2e}
\usepackage{graphicx}
\usepackage{epsfig}
\usepackage{amssymb}
\usepackage{lscape}
\usepackage{ulem}
\usepackage{txfonts}

\def\kms{km~s$^{-1}$}

\def\lya{Ly$\alpha$}

\def\HI{H\,{\sc i}}

\def\c2s{C\,{\sc ii}$^{\star}$}

\def\fgas{$f_{\rm gas}$}
\def\dfgas{$\Delta f_{\rm gas}$}
\def\dfgash2{$\Delta f_{\rm gas, H_2}$}
\def\mhi{M$_{\rm HI}$}

\title[Enhanced gas fractions in recently merged galaxies] {Enhanced atomic gas fractions in recently merged galaxies:  quenching is not a result of post-merger gas exhaustion.}

\author[Ellison et al.] {Sara L. Ellison$^1$, Barbara Catinella$^{2,3}$, Luca Cortese$^{2,3}$ \\ 
$^1$ Department of Physics \& Astronomy, University of Victoria, Finnerty Road, Victoria, British Columbia, 
V8P 1A1, Canada.\\
$^2$ International Centre for Radio Astronomy Research, The University of Western Australia, 35 Stirling Hwy, 
Crawley, WA 6009, Australia\\
$^3$ ARC Centre of Excellence for All Sky Astrophysics in 3 Dimensions (ASTRO 3D)
}
\begin{document}

\maketitle

\begin{abstract}
We present a detailed assessment of the global atomic hydrogen gas fraction (\fgas=log[\mhi/M$_{\star}$])
in a sample of post-merger galaxies identified in the Sloan Digital Sky Survey (SDSS).
Archival \HI\ measurements of 47 targets are combined with new Arecibo observations
of a further 51 galaxies.  The stellar mass range of the post-merger sample, our
observing strategy, detection thresholds and data analysis procedures replicate 
those of the extended GALEX Arecibo SDSS Survey (xGASS) 
which can therefore be used as a control sample.  Our principal results are:
1)  The post-merger sample shows a $\sim$ 50 per cent higher \HI\ detection fraction compared with
xGASS; 2)  Accounting for non-detections, the median atomic gas fraction of the 
post-merger sample is larger than the
control sample by 0.3 -- 0.6 dex;  3)  The median atomic gas fraction enhancement
(\dfgas), computed on a galaxy-by-galaxy basis at fixed stellar mass, is 0.51
dex.    Our results demonstrate that recently
merged galaxies are typically a factor of $\sim$ 3 more \HI\ rich than control
galaxies of the same M$_{\star}$.    If the control sample is additionally matched
in star formation rate, the median \HI\ excess is reduced to \dfgas\ = 0.2 dex,
showing that the enhanced atomic
gas fractions in post-mergers are not purely a reflection of changes in star formation activity.
We conclude that merger-induced starbursts and outflows do not lead to prompt quenching via
exhaustion/expulsion of the galactic gas reservoirs.  Instead, we propose that if star formation
ceases after a merger, it is more likely due to an enhanced turbulence which renders
the galaxy unable to effectively form new stars.

\end{abstract}


\section{Introduction}

It is now well established that galaxy-galaxy interactions can
have a profound effect on both the gas and stellar content of
their participants.  Boosts in star formation, enhanced nuclear
accretion, morphological distortions and re-distribution of metals are both predicted by
simulations (e.g. Di Matteo et al. 2007; Cox et al. 2008; Perez, Michel-Dansac \& Tissera 2011;
Torrey et al. 2012; Moreno et al. 2015; Capelo et al. 2015; Bustamante et al. 2018;
Blecha et al. 2018) and 
confirmed by observations (e.g. Woods \& Geller 2007; Ellison
et al. 2008, 2011, 2013; Scudder et al. 2012; Satyapal et al. 2014; Patton et al. 
2016; Weston et al. 2017; Goulding et al. 2018).  However, much less
is known about the detailed gas physics of interacting galaxies, and
how the budget of the different gas phases evolves during the merger, with limited
insight from simulations and obsevations alike.

Perhaps the most well known effect on the galactic gas content
studied in merger simulations is the central concentration
that results from tidal torques (Mihos \& Hernquist 1994; 1996).
As a result of the internal asymmetries and instabilities caused by
the gravitational interaction, gas inflows dilute the central gas-phase metallicity
(Montuori et al. 2010; Perez et al. 2011;
Rupke et al. 2010; Torrey et al. 2012; Bustamante et al. 2018).
Moreover, Hani et al. (2018) have recently used a zoom-in simulation
of an approximately equal mass merger to show that interactions
can additionally impact their \textit{outer} gas reservoirs.  By
considering the circumgalactic medium (CGM) out to 
$\sim$ 150 kpc, Hani et al.
(2018) find a significant increase in the hydrogen covering
fraction, characteristic gas halo size and metallicity that
can endure several Gyrs after the merger is complete.  
The emerging picture from these merger simulations is
thus that interactions can both funnel gas inwards, and
re-distribute material into the CGM.

Despite the progress of these theoretical works, relatively
few studies have attempted to compile a detailed accounting
of the multi-phase gas budget throughout the merger process.  
Although a few cosmological simulations have found an enhanced
\HI\ fraction in either paired (Tonnesen \& Cen 2012) or
post-merger (Rafieferantsoa et al. 2015) galaxies, these 
simulations lack the spatial resolution to trace the detailed
physics of the interstellar medium (ISM).  The technical
challenges in such an endevour are significant, both in the
implementation of the complex physical processes occuring in
the ISM and also resolving the scales on
which these processes occur (e.g. Teyssier et al. 2010;
Renaud et al. 2014).  

Most previous high resolution simulations have been run on
single interactions, which precludes a statistical study.
Moreno et al. (2018) have recently
made progress in this regard by simulating a suite of galaxy
mergers that incorporate the state-of-the-art Feedback In
Realistic Environments 2 (FIRE-2) code (Hopkins et al. 2018).  These simulations
are run at a resolution sufficient to resolve the processes
of both star formation (at the level of giant molecular clouds)
and its energy feedback\footnote{For reference, the FIRE-2 merger
simulations have mass resolution $\sim 1.5 \times 10^4$ M$_{\odot}$,
a minimum softening of 10 pc and a minimum temperature of 10K. See 
Moreno et al. (2018) for more details.}.  Moreno et al.
(2018) have shown that the balance of gas in different phases
(ionized, atomic and molecular) undergoes a complex exchange
through different times of the merger.  Nonetheless, they conclude that the
net result of these reactions is a boost in the molecular gas
mass in the interaction phase, that is predominantly fuelled
from the atomic gas reservoir.  However, the atomic reservoir
can be simultaneously replenished from gas that cools from
the ionized phase.  Over most of the interaction the combined
effect of conversion of atomic gas into molecular
and cooling of ionized gas into the atomic phase balance out,
leading to only a very small increase ($\sim$ 5 per cent)
in the \HI\ mass (averaged across the simulation suite).

On the observational side, numerous works have mapped the atomic
(e.g. Hibbard \& Yun 1999; Koribalski \& Dickey 2004;
Manthey et al. 2008a, b; Fernandez et al.
2010) or molecular (e.g. Tacconi et al. 1999;
Aalto et al., 2001; Yun \& Hibbard 2001; Wang et al. 2004) gas
content of individual merging
galaxies.  Such studies regularly find that gas is re-distributed
during the merger, which results in both tidal tails and
central molecular gas concentations (e.g. Hibbard \& van Gorkom 1996;
Georgakakis et al. 2000; Ueda et al. 2014; Yamashita et al. 2017).
However, fewer studies have previously attempted to statistically compare
the gas content of mergers (either selected to be close pairs
or single coalesced post-merger galaxies) with that of the non-interacting population.

Investigations of the statistical gas content of mergers
have generally supported the predicted boost in the molecular gas 
content of galaxy mergers  (e.g. Braine \& Combes 1993; 
Combes et al. 1994; Casasola, Bettoni \& Galletta 2004; Larson et al. 2016), 
but also found shorter molecular gas depletion times (molecular gas mass/SFR), 
due to the enhanced SFRs (e.g. Solomon \& Sage 1988; Georgakakis et al. 2000;
Saintonge et al. 2012; Michiyama et al. 2016).
However, these studies are potentially undermined by various technical
limitations, such as sample size, lack of a well matched control sample and
assumptions of single or bimodal CO-to-H$_2$ conversion factors in interacting
and isolated galaxies.  In a pair of companion papers, Violino
et al. (2018) and Sargent et al. (in prep) study the molecular
gas fraction in pairs and post-mergers respectively, with physically 
motivated (continuous) conversion factors and a carefully matched control
sample whose H$_2$ masses were computed with identical procedures, e.g.
aperture corrections and conversion factors.  After these rigourous experimental
procedures, the works of Violino et al. (2018) and Sargent et al. (in prep)
robustly confirm the enhanced molecular gas fractions in pairs (by 0.4 dex) and post-mergers
(by 0.6 dex), respectively (at fixed stellar mass). 
The observed H$_2$ enhancement is in qualitative agreement with the
simulations of Moreno et al. (2018), although the 27 orbital variations of a
1:2.5 mass merger yield a more modest median molecular gas fraction enhancement of
$\sim$ 0.1 dex.

However, the situation for the experimental measurement of the atomic
gas content remains incomplete and more contentious.  Again, although studies of the
atomic gas content of mergers (pairs or post-mergers) have been performed, experimental
problems remain due to either small sample size, lack of a
robust control sample, accounting for multiple galaxies in a
large telescope beam or shallow survey depth.  As a result,
apparently conflicting results have been reported, with some
studies finding little difference between the \HI\ gas fraction (log \fgas\
= log [\mhi/M$_{\star}$]
\footnote{Although we will also discuss molecular
gas fractions in this paper, \fgas\ refers to the atomic gas fraction
unless otherwise stated.})
in mergers and non-mergers (e.g. Braine \& Combes 1993;  Ellison et al.
2015; Stierwalt et al. 2015; Zuo et al. 2018) and others claiming an 
enhanced (Casasola et al. 2004; Huchtmeier et al. 2008; Jaskot et al. 2015), 
or decreased atomic content (Hibbard \& van Gorkom 1996).

In the work presented here, we attempt to tackle several of the
short-comings encountered by previous studies of the \HI\
gas content of galaxy mergers and resolve the controversy of their
atomic gas fractions.  First, we identify
a sample of galaxies in the post-coalescence phase (hereafter,
post-mergers) so that it is not necessary to account for the
combined contribution of two galaxies' 21 cm emission in the large Arecibo beam.
Second, we significantly enlarge previous observational samples -
our final sample consists of observations for 98 post-mergers,
compared to the statistical sample of 37 in our previous work
(Ellison et al. 2015).  Finally, and potentially most importantly,
the study presented here is sensitive to much lower atomic gas fractions
than presented in Ellison et al (2015).  The depth of our earlier
work was dictated by the comparison of \HI\ gas fractions in mergers with
those of a control sample taken from the Arecibo Legacy Fast
ALFA (ALFALFA) survey (Giovanelli et al. 2005).  However, since ALFALFA is a
relatively shallow 21 cm survey, it provides 21 cm detections only
for the most \HI\ gas-rich galaxies at a given stellar mass.  In the
work presented here, our observations are designed to achieve
the much deeper 21 cm sensitivity of the GALEX Arecibo SDSS Survey (GASS;
Catinella et al. 2010, 2013) and its low mass extension, which together make up 
the extended GASS sample (xGASS;
Catinella et al. 2018).  The xGASS survey contains $\sim$ 1200 galaxies
with measurements of \fgas\ that are sensitive to
a few per cent over a mass range 9.0 $<$ log M$_{\star} <$ 11.5.
The xGASS survey therefore represents an excellent control sample,
being both sensitive, large and spanning a broad range of stellar masses.

The paper is laid out as follows.  In Sec. \ref{sample_sec} we
describe our survey strategy and the identification of post-mergers.
Observations performed at the Arecibo telescope, and their data
reduction, are described in Sec. \ref{obs_sec}.  Our results
are presented and discussed in Sec \ref{results_sec} and
\ref{discussion_sec} respectively, and our conclusions are summarized
in Sec. \ref{conclusions_sec}.  We adopt a cosmology in which
H$_0$=70 km/s/Mpc, $\Omega_M$=0.3, $\Omega_\Lambda$=0.7.

\section{Survey strategy and sample selection}\label{sample_sec}

The principal defining strategy of our sample selection
and observations is to follow the procedures of the xGASS
survey (Catinella et al. 2018).  The \HI\ detection threshold
is designed to replicate that of xGASS and the data reduction
and analysis procedures are identical (see Sec. \ref{obs_sec} for
more details).  The sample of post-mergers
selected for this work is therefore chosen to simultaneously
take advantage of extant archival HI measurements but also permit an unbiased
comparison with xGASS.

In order to select our sample of post-mergers, we begin by compiling 
candidates from three sources.  The first is the post-merger catalog
compiled by Darg et al. (2010) from Galaxy Zoo with further visual
classification performed by Ellison et al. (2013) to remove suspect
cases.  The final Galaxy Zoo sample contains 100 post-mergers (see Ellison
et al. 2013 for more details).
To this we add a further 113 galaxies based on the visual classifications
of Nair \& Abraham (2010); any duplicates with the Galaxy Zoo
sample are removed.  We apply a redshift ($z<0.04$) and declination 
($0 < \delta\ < 37$ deg) cut to both the Galaxy Zoo and Nair \& Abraham (2010) samples
in order to define a target list that is readily observable from Arecibo, which
reduces the sample from 213 to only 37.

In order to increase the sample size, we turn to the quantitative morphological
classifications of Simard et al. (2011).  All galaxies within the Arecibo
declination and redshift cut defined above, and with Simard et al. (2011) $r$-band asymmetries $>$ 0.05 
are visually inspected, which yields a further 114 post-merger galaxies.  The
post-merger parent sample is therefore a combination of the
Galaxy Zoo, Nair \& Abraham (2010) and asymmetry-selected samples and
contains 151 post-mergers, all of which should be readily observable
from Arecibo, based on cuts of redshift and declination.

Three further cuts are made to the combined post-merger sample in order to
achieve an unbiased survey design that can be compared with a control
sample selected from xGASS.  First, SDSS images of galaxies in
the post-merger sample are visually checked for companions within 1 arcmin
of the primary target, in order to avoid potential contamination in
the Arecibo beam.  Second, since we will assess the gas fraction in post-mergers
via a quantitative comparison with the xGASS survey,
whose survey limit is log M$_{\star} \ge$ 9.0 M$_{\odot}$, we remove any post-mergers
less massive than this threshold.  Finally, we set a survey footprint that
is bounded not only by Arecibo observing limits, but also by our observing
time allocation (Sec. \ref{obs_sec}).  Thus, in addition to the declination limits described above,
we also require that the right ascension lies in the range 10 $< \alpha\ <$ 17 
hours\footnote{The vast majority of the post-merger sample lies in this
range to start with: only 9 galaxies are removed in the right ascension cut.}.  

\medskip\

In summary, the final post-merger sample is defined by cuts in stellar mass 
(log M$_{\star} \ge$ 9.0 M$_{\odot}$), declination ($0 < \delta < 37$ deg), right
ascension (10 $< \alpha <$ 17 hours) and redshift ($z<0.04$) and contains 107
targets.  

\medskip\

Archival observations exist for 47/107 of the post-mergers in our final sample.
36 \HI\ detections are found in the ALFALFA survey (Haynes et al. 2011 and
the $\alpha$70 data
release\footnote{http://egg.astro.cornell.edu/alfalfa/data/index.php}), 6 detections
from Springob et al. (2005), two detections in Ellison et al. (2015) 
and three galaxies were targetted as part of the xGASS sample, of which two are 
detections and one yields an upper limit.  Since our survey strategy 
is to replicate the depth of the xGASS survey, the xGASS
non-detection was not re-observed.  We do not attempt to expand
the post-merger sample by adding archival observations outside of
our defined sample.  Doing so has the potential to bias gas
fractions if previous surveys have not followed the same consistent
depth and survey strategy.

\section{Arecibo observations}\label{obs_sec}

\begin{figure*}
\begin{center}
\includegraphics[width=17cm]{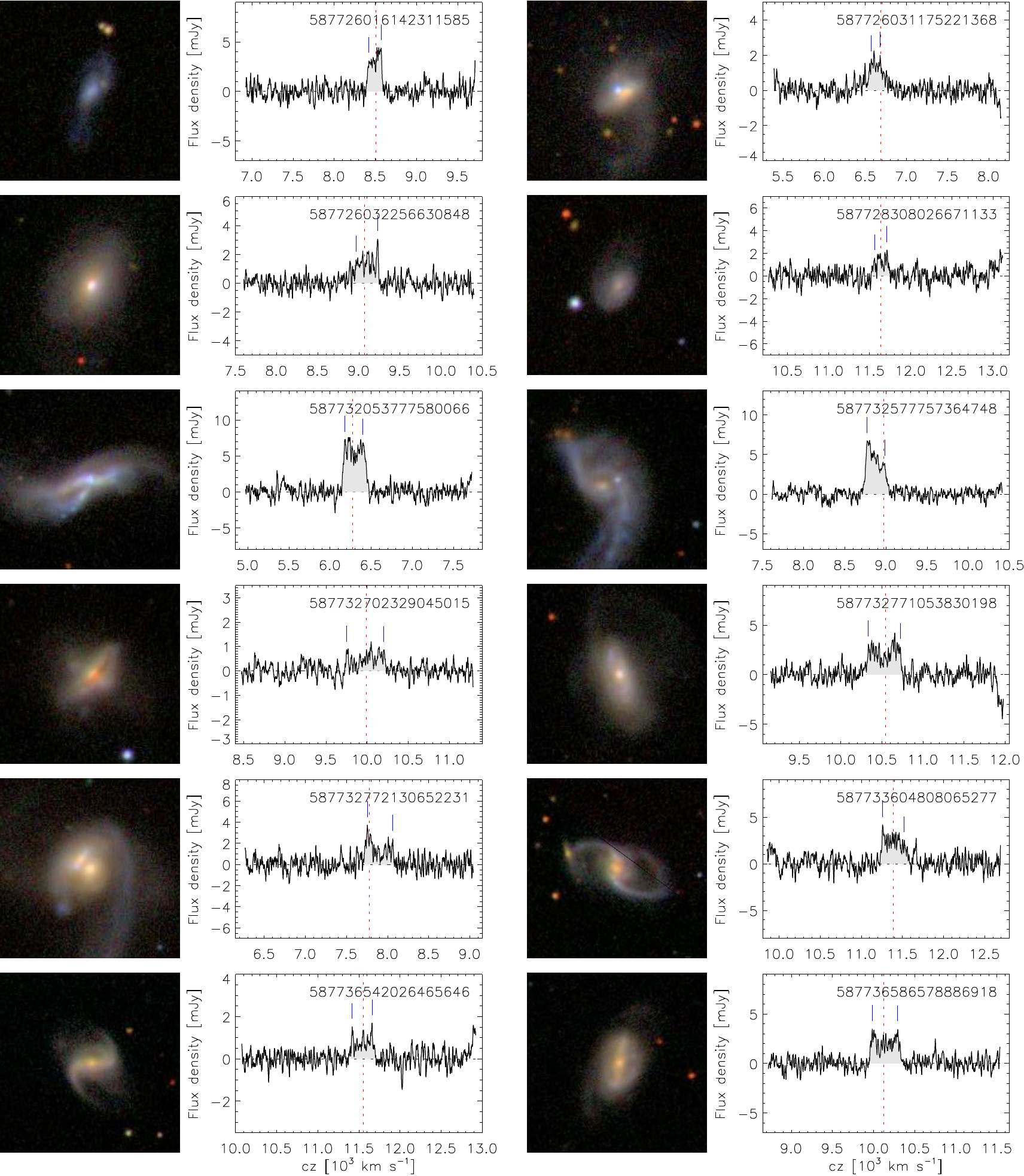}
\caption{SDSS postage stamp images (1 arcmin square) and \HI -line profiles of the galaxies detected in this work 
for the first 12 galaxies in Table \ref{obs_table}, ordered by SDSS objID (as indicated in the panels). 
The \HI\ spectra are calibrated, smoothed and baseline-subtracted. A dotted line and two dashes indicate the 
heliocentric velocity corresponding to the SDSS redshift and the two peaks used for width measurement,
respectively.}
\label{detect_fig}
\end{center}
\end{figure*}

Accounting for the 47 previously observed post-mergers leaves 60 additional targets
requiring new observations.  Since the sample is selected within the
right ascension criterion 10 $< \alpha <$ 17 hours, all unobserved
galaxies could be targeted with a single 32 hour Arecibo proposal 
concentrated on the Spring sky, with observation blocks spread between
January and April 2017.

Our objective was to replicate the observing strategy of the xGASS survey, sampling
galaxies randomly selected from the SDSS down to a stellar mass limit of
log M$_{\star}$ = 9.0 M$_{\odot}$.  The GASS survey has an atomic
gas fraction limit of 1.5 percent at log M$_{\star} >$ 10.5 M$_{\odot}$
and reaches a fixed log \mhi\ = 8.7 at stellar masses down to log M$_{\star} =$ 10.0
(Catinella et al. 2010).  The low mass extension to GASS also
aims to achieve atomic gas fraction detections down to a few per cent,
although with a slightly different set of mass dependent  criteria
(see Catinella et al. 2018).  The design of our post-merger observations aimed to achieve
similar depths to xGASS such that detection fractions and average
atomic gas fractions could be fairly compared.

In practice, 51/60 of the previously unobserved post-mergers in our
sample were observed, with insufficient time to complete the final
9 targets.  The 9 unobserved targets represent a random subset.

\medskip

The Arecibo observations were carried out in six sessions, between January and April 2017,
with 32 hours allocated under program A3117 (PI Ellison).
We observed remotely in standard position-switching mode, using the 
L-band wide receiver and the interim correlator as a backend. Two correlator boards with 
12.5 MHz bandwidth, one polarization, and 2048 channels per spectrum (yielding
a velocity resolution of 1.4 \kms\ at 1370 MHz before smoothing) were centered
at or near the frequency corresponding to the SDSS redshift of the target. 
The \HI\ spectra were recorded every second with 9-level sampling.

We observed each galaxy until detected with a signal-to-noise of 5 or higher, or until
a gas fraction limit of $\sim$2\% was reached. Consistently with xGASS, upper limits for 
non-detections are computed assuming a 5$\sigma$ signal with 300 \kms\
velocity width (appropriate since all of the non-detections are for galaxies
with log M$_{\star} >$ 10 M$_{\odot}$), if the spectrum was smoothed to half width.

The data reduction, performed in the IDL environment, includes
Hanning smoothing, bandpass subtraction, excision of radio frequency interference (RFI), and flux
calibration. The spectra obtained from each on/off pair are weighted by 1/$rms^2$,
where $rms$ is the root mean square noise measured in the
signal-free portion of the spectrum, and co-added. The two
orthogonal linear polarizations (kept separated up to this point) are
averaged to produce the final spectrum. The spectrum is then boxcar smoothed,
baseline subtracted and measured as explained in Catinella et al. (2010). 
Velocity widths are measured at the 50\% of each flux peak, and corrected
for instrumental broadening and cosmological redshift as described in
Catinella et al. (2012). The RFI excision technique is illustrated in detail in 
Catinella \& Cortese (2015).

\medskip

A summary of observations for the 51 newly observed post-mergers is given
in Table \ref{obs_table}.  The upper part of the table reports the 43
detections and the lower part lists the 8 non-detections.  Fig. \ref{detect_fig}
shows the first 12 detections listed in Table \ref{obs_table}, which are ordered
by SDSS objID.  For each
galaxy we show the SDSS postage stamp and the extracted spectrum.
Panels showing the remaining detections, as well as the non-detections,
are presented in the Appendix.

\HI\ masses (in solar units) are computed via the redshift dependent
equation (e.g. Catinella et al. 2018):

\begin{equation}
\frac{M_{HI}}{M_{\odot}} = \frac{2.356 \times 10^5}{(1+z_{HI})^2} \bigg[\frac{d_L(z)}{Mpc}\bigg]^2 \bigg( \frac{\int S dv}{Jy~km~s^{-1}} \bigg)
\end{equation}

where $d_L(z)$ is the luminosity distance to the galaxy at
redshift $z_{HI}$ as measured from the \HI\ spectrum (column 4 in Table \ref{obs_table}).
For non-detections, the upper limit for the \HI\ mass was computed
using the SDSS redshift  (column 4 in Table \ref{results_table}).
Table \ref{results_table} lists the compiled \HI\ masses and upper limits
for the 107 post-mergers in our final sample - 47 values from the literature,
51 new observations and 9 unobserved targets.

\begin{table*}
\begin{center}
\caption{Arecibo observations of post-mergers, ordered by SDSS objID.}
\begin{tabular}{llrccccccc}
\hline\\
objID                &             RA      &            Dec      &        z$_{HI}$ &       T$_{on}$ &          W50   &       $S_{obs}$ &       $S_{err}$ &            RMS  &  S/N\\  
                     &            (deg)     &          (deg)      &               &         (min) &         (km/s)  &      (Jy km/s) &      (Jy km/s) &           (mJy)  &     \\  
\hline  \\  
587726016142311585   &           179.23955 &             3.28063 &       0.028325 &              8 &          167.6 &           0.59 &           0.08 &            0.69 &      12.9 \\ 
587726031175221368   &           180.99823 &             1.41085 &       0.022101 &             30 &          159.4 &           0.27 &           0.04 &            0.35 &      11.8 \\ 
587726032256630848   &           198.46830 &             2.13257 &       0.030343 &             19 &          268.1 &           0.41 &           0.06 &            0.43 &      11.3 \\ 
587728308026671133   &           164.08866 &             1.76512 &       0.038801 &             10 &          170.8 &           0.26 &           0.07 &            0.59 &       6.3 \\ 
587732053777580066   &           127.22636 &            34.65166 &       0.021013 &              5 &          266.8 &           1.68 &           0.10 &            0.75 &      27.1 \\ 
587732577757364748   &           124.47457 &             3.37657 &       0.029620 &             10 &          245.9 &           1.23 &           0.08 &            0.59 &      25.6 \\ 
587732702329045015   &           160.29066 &             6.82006 &       0.033293 &             35 &          454.5 &           0.23 &           0.05 &            0.27 &       6.5 \\ 
587732771053830198   &           178.60401 &             9.60905 &       0.035119 &             10 &          387.7 &           0.91 &           0.11 &            0.67 &      13.3 \\ 
587732772130652231   &           185.74051 &            10.54832 &       0.026352 &             10 &          322.5 &           0.53 &           0.10 &            0.67 &       8.6 \\ 
587733604808065277   &           253.34229 &            33.02841 &       0.038028 &              5 &          286.1 &           0.76 &           0.11 &            0.74 &      11.6 \\ 
587736542026465646   &           233.99535 &             6.38257 &       0.038487 &             25 &          241.5 &           0.23 &           0.05 &            0.35 &       8.0 \\ 
587736586578886918   &           225.20738 &            36.03508 &       0.033821 &             10 &          331.8 &           0.78 &           0.09 &            0.59 &      14.1 \\ 
587736619864817968   &           247.13706 &            23.59831 &       0.038373 &              4 &          196.8 &           0.83 &           0.10 &            0.81 &      13.9 \\ 
587736941982777562   &           217.83852 &            35.62303 &       0.034767 &              4 &          466.7 &           2.74 &           0.18 &            0.96 &      23.1 \\ 
587738952029831251   &           201.08403 &            36.59609 &       0.015834 &              5 &          251.9 &           3.28 &           0.12 &            0.90 &      45.6 \\ 
587738952029962307   &           201.40139 &            36.38123 &       0.018771 &              5 &          102.9 &           2.26 &           0.07 &            0.91 &      54.0 \\ 
587739294552358946   &           169.65756 &            33.48587 &       0.037503 &             10 &          198.0 &           0.78 &           0.07 &            0.61 &      17.5 \\ 
587739305294168109   &           194.56861 &            35.83894 &       0.037310 &             24 &          340.5 &           0.52 &           0.07 &            0.41 &      13.3 \\ 
587739376693936354   &           125.18686 &            19.36220 &       0.019187 &              5 &          312.7 &           1.95 &           0.11 &            0.75 &      29.0 \\ 
587739608639209570   &           215.92608 &            28.34609 &       0.030889 &              9 &          508.4 &           1.22 &           0.12 &            0.62 &      14.8 \\ 
587739609692307530   &           161.12063 &            30.66089 &       0.034774 &              5 &          242.0 &           1.56 &           0.10 &            0.73 &      26.5 \\ 
587739646205624406   &           160.75167 &            29.10123 &       0.035487 &              8 &          209.2 &           0.71 &           0.07 &            0.59 &      16.0 \\ 
587739651571384503   &           239.38416 &            20.05347 &       0.032707 &              5 &          272.5 &           1.72 &           0.12 &            0.84 &      24.0 \\ 
587739652642373881   &           233.54299 &            23.49964 &       0.035863 &              5 &           91.6 &           0.43 &           0.06 &            0.69 &      12.1 \\ 
587739827667402818   &           221.18423 &            22.43805 &       0.032527 &              9 &          251.9 &           0.29 &           0.08 &            0.47 &       5.8 \\ 
587739828200210569   &           211.44225 &            25.16498 &       0.031591 &              5 &           50.9 &           0.22 &           0.05 &            0.80 &       7.0 \\ 
587741709954121847   &           168.41890 &            27.24117 &       0.037079 &             20 &          209.3 &           0.30 &           0.05 &            0.37 &      10.6 \\ 
587741819989655635   &           144.87699 &            19.12943 &       0.026500 &              9 &          167.2 &           0.65 &           0.07 &            0.66 &      14.7 \\ 
587741830195904672   &           169.65079 &            24.86748 &       0.025086 &             15 &          250.8 &           0.40 &           0.07 &            0.52 &       9.6 \\ 
587742008968675684   &           120.76626 &             9.67965 &       0.034242 &             13 &          187.3 &           0.60 &           0.07 &            0.55 &      15.4 \\ 
587742189913833708   &           206.55437 &            23.08402 &       0.030016 &              5 &          186.9 &           0.82 &           0.08 &            0.71 &      16.3 \\ 
587742191514878026   &           182.64450 &            25.92821 &       0.021446 &              4 &          237.2 &           0.59 &           0.11 &            0.81 &       9.3 \\ 
587742577534500950   &           229.06901 &            15.27590 &       0.036486 &              9 &          180.4 &           0.49 &           0.07 &            0.56 &      12.4 \\ 
588010358543941775   &           167.65603 &             4.19140 &       0.028927 &              4 &          130.9 &           0.83 &           0.09 &            0.88 &      15.7 \\ 
588010359073931336   &           151.83607 &             4.07930 &       0.028670 &              4 &          231.2 &           0.86 &           0.11 &            0.83 &      13.3 \\ 
588016878824063265   &           120.11882 &            21.13626 &       0.030256 &             10 &          259.0 &           1.17 &           0.08 &            0.54 &      26.1 \\ 
588017566018240608   &           165.28135 &            11.37236 &       0.032988 &              5 &           97.7 &           0.34 &           0.08 &            0.93 &       7.0 \\ 
588017603627909293   &           233.83991 &            32.21680 &       0.036686 &              5 &          254.8 &           1.24 &           0.11 &            0.80 &      18.6 \\ 
588017702952370465   &           230.72835 &             8.34895 &       0.031184 &             20 &          175.1 &           0.39 &           0.04 &            0.39 &      14.6 \\ 
588017977842008451   &           243.55236 &            22.77767 &       0.038992 &              7 &          166.2 &           0.57 &           0.07 &            0.66 &      12.8 \\ 
588017978365247555   &           211.19749 &            36.60761 &       0.034756 &              9 &          182.7 &           0.61 &           0.08 &            0.67 &      13.1 \\ 
588018091085791271   &           250.59864 &            25.08653 &       0.022491 &              4 &          392.0 &           4.93 &           0.14 &            0.92 &      59.6 \\ 
588297863638941792   &           128.88771 &            30.53429 &       0.025558 &              5 &          178.7 &           2.29 &           0.13 &            1.06 &      31.5 \\ 
\hline  \\  
587726032235921453   &           151.13947 &             1.78775 &     ...        &             19 &     ...        &     ...        &     ...        &            0.43 &     ...  \\ 
587729160042119249   &           199.24912 &             5.65679 &     ...        &             10 &     ...        &     ...        &     ...        &            0.43 &     ...  \\ 
587736584973844652   &           237.57736 &            28.81188 &     ...        &              8 &     ...        &     ...        &     ...        &            0.52 &     ...  \\ 
587739647816761430   &           161.79663 &            30.72437 &     ...        &              4 &     ...        &     ...        &     ...        &            0.72 &     ...  \\ 
587739719755300871   &           198.10772 &            28.53771 &     ...        &              8 &     ...        &     ...        &     ...        &            0.57 &     ...  \\ 
587742628523475193   &           219.30721 &            14.66514 &     ...        &             16 &     ...        &     ...        &     ...        &            0.41 &     ...  \\ 
588017703491469448   &           235.86262 &             7.91169 &     ...        &             15 &     ...        &     ...        &     ...        &            0.37 &     ...  \\ 
588023668630618202   &           173.68616 &            19.03339 &     ...        &              8 &     ...        &     ...        &     ...        &            0.68 &     ...  \\ 
\hline  \\  
\end{tabular}
\label{obs_table}
\end{center}
\end{table*}

\begin{table*}
\begin{center}
\caption{Sample properties and atomic gas fractions of the final post-merger sample, ordered by SDSS objID.  \dfgas\ calculations include
non-detections and are computed for galaxies with log M$_{\star} \le 10.8$ M$_{\odot}$.}
\begin{tabular}{lccccccrc}
\hline\\
SDSS objID            &     RA &      Dec   & z$_{SDSS}$ &    log M$_{\star}$ & log M$_{HI}$ &     log f$_{gas}$ &   $\Delta$ f$_{gas}$ &  HI Reference \\  
                     &  (deg) &    (deg)   &           & (M$_{\odot}$)  &  (M$_{\odot}$)  &     (dex)       &            (dex)         &                 \\  
\hline  \\  
587726016142311585   &           179.23955 &             3.28063 &         0.0284 &       9.16 &        9.31 &         0.15 &     0.29     &     This work            \\ 
587726031175221368   &           180.99823 &             1.41085 &         0.0223 &      10.11 &        8.75 &        -1.36 &     -0.32    &     This work            \\ 
587726032235921453   &            151.13947 &             1.78775 &          0.0310 &      10.60 &     $<$8.93 &     $<$-1.67 &     $<$-0.37  &     This work            \\ 
587726032256630848   &           198.46830 &             2.13257 &         0.0303 &      10.53 &        9.21 &        -1.32 &     0.08     &     This work            \\ 
587728308026671133   &           164.08866 &             1.76512 &         0.0388 &       9.86 &        9.21 &        -0.65 &     0.05     &     This work            \\ 
587728309101789218   &           167.22810 &             2.67660 &         0.0357 &      10.34 &        9.70 &        -0.64 &     0.72     &     ALFALFA              \\ 
587728879262105824   &           144.02909 &             3.01169 &         0.0187 &       9.22 &        9.48 &         0.26 &     0.45     &     ALFALFA              \\ 
587728879794192715   &           133.04220 &             2.84024 &         0.0290 &      10.31 &        9.94 &        -0.37 &     0.84     &     ALFALFA              \\ 
587728949589377044   &            161.26500 &             0.43439 &         0.0263 &      10.29 &     ...     &     ...      &     ...      &     ...                  \\ 
587729158966345769   &           194.58816 &             4.88565 &         0.0361 &      10.37 &        9.98 &        -0.39 &     1.05     &     ALFALFA              \\ 
587729160042119249   &            199.24912 &             5.65679 &          0.0331 &      10.74 &     $<$8.99 &     $<$-1.75 &     $<$-0.45  &     This work            \\ 
587732053777580066   &           127.22636 &            34.65166 &         0.0209 &       9.87 &        9.50 &        -0.37 &     0.43     &     This work            \\ 
587732577757364748   &           124.47457 &             3.37657 &         0.0299 &       9.69 &        9.66 &        -0.03 &     0.70     &     This work            \\ 
587732702329045015   &           160.29066 &             6.82006 &         0.0333 &      10.59 &        9.04 &        -1.55 &     -0.25    &     This work            \\ 
587732771053830198   &           178.60401 &             9.60905 &         0.0352 &      10.66 &        9.68 &        -0.98 &     0.25     &     This work            \\ 
587732772130652231   &           185.74051 &            10.54832 &         0.0259 &      10.60 &        9.19 &        -1.41 &     -0.11    &     This work            \\ 
587733411530932307   &            243.64022 &            36.94379 &         0.0380 &      10.32 &     ...     &     ...      &     ...      &     ...                  \\ 
587733604808065277   &           253.34229 &            33.02841 &         0.0380 &      10.54 &        9.67 &        -0.87 &     0.51     &     This work            \\ 
587734622163763258   &           120.86694 &            25.10267 &         0.0276 &      10.13 &        9.18 &        -0.95 &     0.07     &     Springob et al. (2005) \\ 
587734893287440431   &           174.04155 &            10.05556 &         0.0207 &      10.37 &        9.73 &        -0.64 &     0.80     &     ALFALFA              \\ 
587736478124408901   &            209.25492 &            11.61143 &         0.0387 &      10.80 &     ...     &     ...      &     ...      &     ...                  \\ 
587736542026465646   &           233.99535 &             6.38257 &         0.0386 &      10.43 &        9.15 &        -1.28 &     0.27     &     This work            \\ 
587736543096799321   &           226.32453 &             8.15355 &         0.0391 &      10.11 &        9.81 &        -0.30 &     0.74     &     ALFALFA              \\ 
587736546311340046   &           225.62214 &             5.54494 &         0.0376 &      10.62 &       10.40 &        -0.22 &     1.03     &     ALFALFA              \\ 
587736584973844652   &           237.57736 &            28.81188 &         0.0308 &      10.95 &     $<$9.01 &     $<$-1.94 &     ...      &     This work            \\ 
587736585511305375   &           239.12874 &            28.43732 &         0.0214 &       9.25 &        9.40 &         0.15 &     0.34     &     ALFALFA              \\ 
587736586578886918   &           225.20738 &            36.03508 &         0.0338 &      10.48 &        9.58 &        -0.90 &     0.63     &     This work            \\ 
587736619864817968   &           247.13706 &            23.59831 &         0.0384 &       9.60 &        9.71 &         0.11 &     0.67     &     This work            \\ 
587736808838594663   &            207.99749 &            13.96750 &         0.0367 &      11.27 &     ...     &     ...      &     ...      &     ...                  \\ 
587736808840560852   &           212.67228 &            13.55800 &         0.0163 &      10.33 &        8.67 &        -1.66 &     -0.28    &     Springob et al. (2005) \\ 
587736916750499857   &           215.63870 &            13.71719 &         0.0255 &      10.31 &       10.13 &        -0.18 &     1.03     &     ALFALFA              \\ 
587736941982777562   &           217.83852 &            35.62303 &         0.0348 &      10.84 &       10.15 &        -0.69 &     ...      &     This work            \\ 
587738409254453314   &           156.84348 &            12.28372 &         0.0330 &       9.58 &        9.57 &        -0.01 &     0.55     &     ALFALFA              \\ 
587738564942102919   &           121.85340 &             6.86320 &         0.0150 &       9.82 &        9.49 &        -0.33 &     0.39     &     ALFALFA              \\ 
587738617018122300   &           149.71387 &            32.07308 &         0.0269 &      10.69 &        9.94 &        -0.75 &     0.49     &     xGASS                \\ 
587738952029831251   &           201.08403 &            36.59609 &         0.0159 &       9.34 &        9.55 &         0.21 &     0.45     &     This work            \\ 
587738952029962307   &           201.40139 &            36.38123 &         0.0187 &      10.10 &        9.53 &        -0.57 &     0.47     &     This work            \\ 
587739159266590725   &           153.54405 &            34.34299 &         0.0376 &      10.34 &        9.83 &        -0.51 &     0.85     &     Ellison et al. (2015) \\ 
587739294552358946   &           169.65756 &            33.48587 &         0.0377 &      10.09 &        9.67 &        -0.42 &     0.61     &     This work            \\ 
587739305294168109   &           194.56861 &            35.83894 &         0.0374 &      10.57 &        9.49 &        -1.08 &     0.22     &     This work            \\ 
587739376693936354   &           125.18686 &            19.36220 &         0.0195 &       9.59 &        9.49 &        -0.10 &     0.46     &     This work            \\ 
587739382067822837   &           234.00632 &            25.55095 &         0.0362 &      10.43 &        9.88 &        -0.55 &     1.00     &     ALFALFA              \\ 
587739406784790531   &           192.17092 &            34.47761 &         0.0142 &       9.42 &        9.35 &        -0.07 &     0.32     &     xGASS                \\ 
587739407324479578   &            200.14753 &            34.13933 &         0.0232 &      10.29 &     ...     &     ...      &     ...      &     ...                  \\ 
587739507154288785   &           185.06547 &            33.66081 &         0.0216 &      10.70 &        9.62 &        -1.08 &     0.16     &     Ellison et al. (2015) \\ 
587739608082481247   &           163.43365 &            29.97818 &         0.0339 &      10.48 &       10.04 &        -0.44 &     1.09     &     ALFALFA              \\ 
587739608639209570   &           215.92608 &            28.34609 &         0.0306 &      10.80 &        9.70 &        -1.10 &     0.27     &     This work            \\ 
587739609692307530   &           161.12063 &            30.66089 &         0.0348 &      10.21 &        9.90 &        -0.31 &     0.72     &     This work            \\ 
587739646205624406   &           160.75167 &            29.10123 &         0.0355 &      10.21 &        9.58 &        -0.63 &     0.40     &     This work            \\ 
587739647816761430   &           161.79663 &            30.72437 &         0.0343 &      11.10 &     $<$9.24 &     $<$-1.86 &     ...      &     This work            \\ 
587739648351338524   &           155.80313 &            29.86167 &         0.0376 &      10.43 &       10.42 &        -0.01 &     1.54     &     ALFALFA              \\ 
587739651571384503   &           239.38416 &            20.05347 &         0.0327 &       9.89 &        9.89 &         0.00 &     0.83     &     This work            \\ 
587739652642373881   &           233.54299 &            23.49964 &         0.0359 &       9.19 &        9.37 &         0.18 &     0.36     &     This work            \\ 
587739719755300871   &            198.10772 &            28.53771 &          0.0213 &      10.44 &     $<$8.73 &     $<$-1.71 &     $<$-0.21  &     This work            \\ 

\hline  \\  
\end{tabular}
\label{results_table}
\end{center}
\end{table*}

\begin{table*}
\begin{center}
\addtocounter{table}{-1}
\caption{Sample properties and atomic gas fractions of the final post-merger sample - continued.}
\begin{tabular}{lccccccrc}
\hline\\
SDSS objID           &     RA &      Dec   & z$_{SDSS}$ &    log M$_{\star}$ & log M$_{HI}$ &     log f$_{gas}$ &   $\Delta$ f$_{gas}$ &  HI Reference \\  
                     &  (deg) &    (deg)   &           & (M$_{\odot}$)  &  (M$_{\odot}$)  &     (dex)       &            (dex)         &                 \\  
\hline  \\  
587739809956495377   &           228.95128 &            20.02236 &         0.0363 &      10.63 &        9.60 &        -1.03 &     0.19     &     ALFALFA              \\ 
587739827136954541   &           235.88197 &            17.31290 &         0.0301 &      10.51 &        9.89 &        -0.62 &     0.80     &     ALFALFA              \\ 
587739827667402818   &           221.18423 &            22.43805 &         0.0326 &      10.83 &        9.12 &        -1.71 &     ...      &     This work            \\ 
587739828200210569   &           211.44225 &            25.16498 &         0.0316 &       9.32 &        8.96 &        -0.36 &     -0.12    &     This work            \\ 
587739828743962776   &           228.00939 &            21.29817 &         0.0158 &      10.69 &        9.88 &        -0.81 &     0.43     &     Springob et al. (2005) \\ 
587739843773857915   &           219.84399 &            19.99086 &         0.0302 &      11.02 &        9.93 &        -1.09 &     ...      &     ALFALFA              \\ 
587739845393186912   &           240.21455 &            15.15126 &         0.0340 &      10.47 &       10.01 &        -0.46 &     1.10     &     ALFALFA              \\ 
587741386735812754   &            120.49371 &            15.05875 &          0.0161 &      10.10 &     $<$8.45 &     $<$-1.65 &     $<$-0.61  &     xGASS                \\ 
587741490357862526   &           134.36642 &            20.12898 &         0.0313 &       9.93 &        9.70 &        -0.23 &     0.66     &     ALFALFA              \\ 
587741490371166352   &           166.47407 &            28.79972 &         0.0331 &      11.02 &       10.15 &        -0.87 &     ...      &     ALFALFA              \\ 
587741533301768296   &           121.92075 &            14.94325 &         0.0288 &      10.55 &        9.74 &        -0.81 &     0.57     &     ALFALFA              \\ 
587741709943308422   &           141.88987 &            21.51811 &         0.0365 &      10.58 &       10.01 &        -0.57 &     0.73     &     ALFALFA              \\ 
587741709954121847   &           168.41890 &            27.24117 &         0.0371 &      10.16 &        9.24 &        -0.92 &     0.11     &     This work            \\ 
587741721213534291   &           195.87117 &            26.55050 &         0.0221 &      10.57 &        9.94 &        -0.63 &     0.67     &     ALFALFA              \\ 
587741726041309213   &           183.67686 &            24.98057 &         0.0220 &       9.89 &        9.34 &        -0.55 &     0.28     &     ALFALFA              \\ 
587741815711858726   &           157.65187 &            21.85477 &         0.0218 &       9.84 &        9.44 &        -0.40 &     0.26     &     Springob et al. (2005) \\ 
587741819989655635   &           144.87699 &            19.12943 &         0.0266 &       9.80 &        9.29 &        -0.51 &     0.21     &     This work            \\ 
587741830195904672   &           169.65079 &            24.86748 &         0.0251 &      10.39 &        9.03 &        -1.36 &     0.13     &     This work            \\ 
587742008968675684   &           120.76626 &             9.67965 &         0.0342 &      10.31 &        9.48 &        -0.83 &     0.38     &     This work            \\ 
587742060544720960   &           206.10219 &            20.40956 &         0.0270 &      11.15 &       10.10 &        -1.05 &     ...      &     ALFALFA              \\ 
587742062170603854   &           241.82595 &            13.22804 &         0.0353 &       9.22 &        9.85 &         0.63 &     0.82     &     ALFALFA              \\ 
587742189913833708   &           206.55437 &            23.08402 &         0.0301 &       9.36 &        9.50 &         0.14 &     0.43     &     This work            \\ 
587742191514878026   &           182.64450 &            25.92821 &         0.0215 &      10.52 &        9.06 &        -1.46 &     -0.02    &     This work            \\ 
587742577534500950   &           229.06901 &            15.27590 &         0.0365 &      10.35 &        9.44 &        -0.91 &     0.47     &     This work            \\ 
587742610812239923   &           241.15390 &             9.95316 &         0.0340 &       9.76 &        9.76 &         0.00 &     0.68     &     ALFALFA              \\ 
587742611343802434   &           229.25508 &            13.10297 &         0.0295 &       9.79 &        9.95 &         0.16 &     0.93     &     Springob et al. (2005) \\ 
587742628523475193   &           219.30721 &            14.66514 &         0.0379 &      10.88 &     $<$9.08 &     $<$-1.8  &     ...      &     This work            \\ 
587744638026907800   &           124.46932 &            12.89799 &         0.0323 &      10.66 &        9.96 &        -0.70 &     0.53     &     ALFALFA              \\ 
587744874250240260   &            125.74443 &            10.39764 &         0.0010 &      10.67 &     ...     &     ...      &     ...      &     ...                  \\ 
587745243615199518   &           120.98490 &             8.69958 &         0.0167 &       9.76 &        9.90 &         0.14 &     0.82     &     ALFALFA              \\ 
587745243629158413   &            152.65950 &            16.68502 &         0.0170 &       9.56 &     ...     &     ...      &     ...      &     ...                  \\ 
588010358531031086   &           138.01528 &             2.87586 &         0.0256 &      10.30 &        9.70 &        -0.60 &     0.54     &     ALFALFA              \\ 
588010358543941775   &           167.65603 &             4.19140 &         0.0290 &       9.51 &        9.47 &        -0.04 &     0.48     &     This work            \\ 
588010359073931336   &           151.83607 &             4.07930 &         0.0286 &      10.26 &        9.48 &        -0.78 &     0.33     &     This work            \\ 
588010359086579785   &           180.80914 &             4.83215 &         0.0385 &       9.64 &        9.77 &         0.13 &     0.67     &     ALFALFA              \\ 
588010360148656136   &           153.99977 &             4.95476 &         0.0320 &      10.41 &        9.89 &        -0.52 &     0.92     &     ALFALFA              \\ 
588010879833145381   &            180.72790 &             5.61263 &         0.0193 &       9.80 &     ...     &     ...      &     ...      &     ...                  \\ 
588010879841992790   &           201.04017 &             5.25972 &         0.0249 &       9.90 &        9.80 &        -0.10 &     0.83     &     ALFALFA              \\ 
588010880378404942   &           199.99113 &             5.80788 &         0.0213 &      10.31 &        9.76 &        -0.55 &     0.66     &     ALFALFA              \\ 
588016878824063265   &           120.11882 &            21.13626 &         0.0302 &      10.44 &        9.66 &        -0.78 &     0.72     &     This work            \\ 
588017566018240608   &           165.28135 &            11.37236 &         0.0330 &       9.11 &        9.19 &         0.08 &     0.17     &     This work            \\ 
588017569779744827   &            203.10099 &            11.10636 &         0.0315 &      10.03 &     ...     &     ...      &     ...      &     ...                  \\ 
588017603627909293   &           233.83991 &            32.21680 &         0.0368 &       9.69 &        9.85 &         0.16 &     0.89     &     This work            \\ 
588017702952370465   &           230.72835 &             8.34895 &         0.0312 &       9.24 &        9.20 &        -0.04 &     0.15     &     This work            \\ 
588017703491469448   &            235.86262 &             7.91169 &          0.0368 &      10.75 &     $<$9.01 &     $<$-1.74 &     $<$-0.43  &     This work            \\ 
588017977842008451   &           243.55236 &            22.77767 &         0.0389 &       9.86 &        9.57 &        -0.29 &     0.41     &     This work            \\ 
588017978365247555   &           211.19749 &            36.60761 &         0.0348 &       9.05 &        9.50 &         0.45 &     0.53     &     This work            \\ 
588017978916667581   &           245.99499 &            22.39337 &         0.0371 &      10.69 &       10.27 &        -0.42 &     0.82     &     Springob et al. (2005) \\ 
588018091085791271   &           250.59864 &            25.08653 &         0.0226 &      10.01 &       10.03 &         0.02 &     0.97     &     This work            \\ 
588023046943473782   &           144.29931 &            21.66922 &         0.0202 &      10.94 &        9.79 &        -1.15 &     ...      &     ALFALFA              \\ 
588023668630618202   &           173.68616 &            19.03339 &         0.0326 &      11.06 &     $<$9.17 &     $<$-1.89 &     ...      &     This work            \\ 
588297863638941792   &           128.88771 &            30.53429 &         0.0255 &       9.76 &        9.81 &         0.05 &     0.73     &     This work            \\ 
588848900971888657   &           146.79909 &             0.70269 &         0.0306 &       9.56 &       10.17 &         0.61 &     1.19     &     ALFALFA              \\ 

\hline  \\  
\end{tabular}
\end{center}
\end{table*}

\section{Results}\label{results_sec}

\begin{figure}
	\includegraphics[width=\columnwidth]{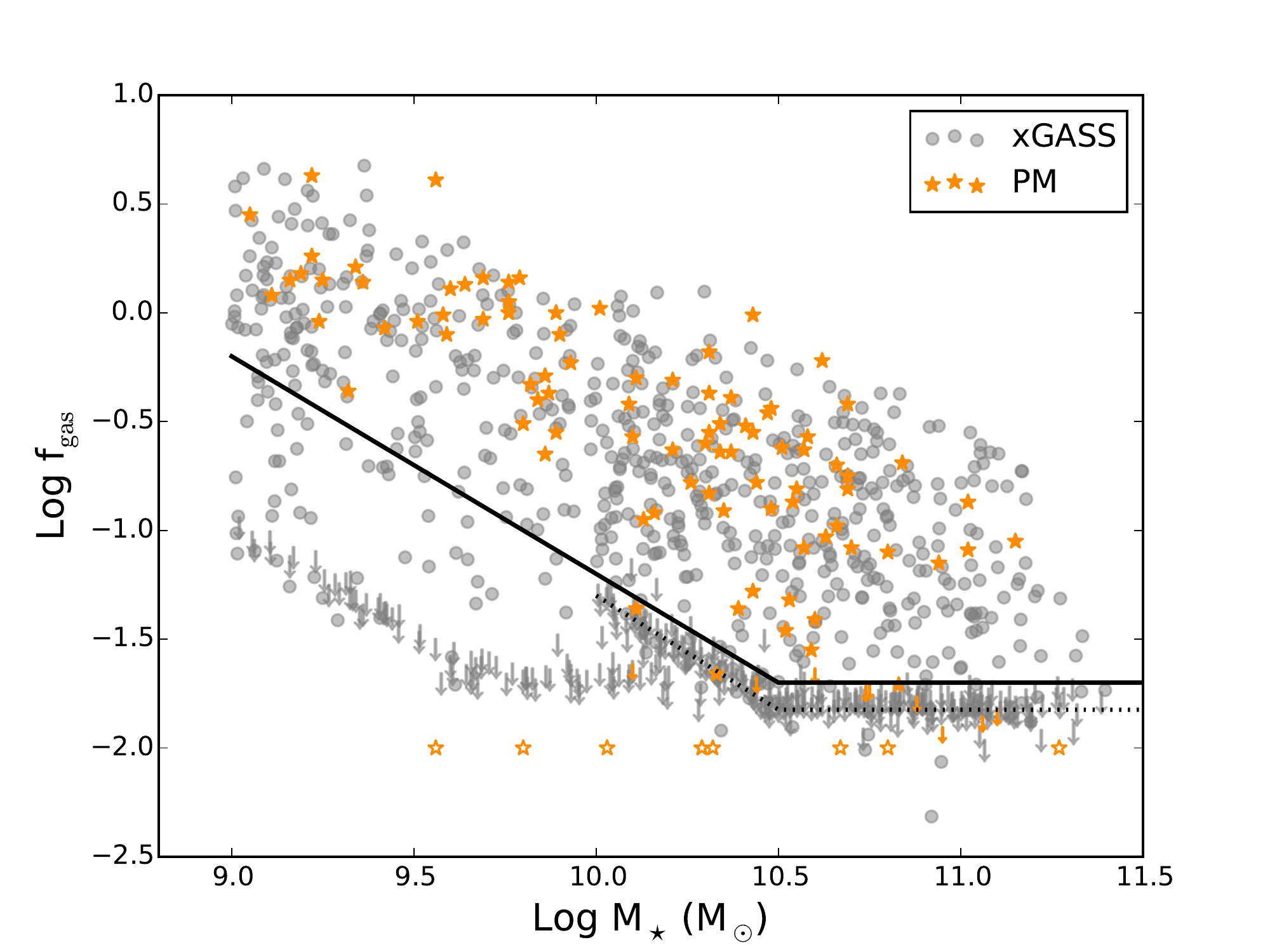}
        \caption{\HI\ gas fractions as a function of stellar mass for post-mergers
          (orange stars) and xGASS (grey circles)  For both samples, upper limits
          are shown by downward pointing arrows.  Open stars show the stellar masses
          of 9 post-mergers in our final sample of 107 that were not observed; their
          gas fractions are arbitrarily shown as log \fgas\ = $-2$ for display purposes.
          The dotted line shows the detection threshold of the GASS survey
          (log [M$_{\star}$/ M$_{\odot}] >$ 10), with
          our modified threshold shown by the solid  line that extends to lower
          masses more conservatively
          demarcates the detections from the upper limits.  The modified detection
          threshold corresponds to  \fgas\ $< $ 2 percent for log (M$_{\star}/M_{\odot}) >$ 10.5
           and log \mhi\ = 8.8  M$_{\odot}$ below that mass.}
    \label{fgas}
\end{figure}

In Fig. \ref{fgas} we show the distribution of \HI\ gas fractions in the post-merger
sample with orange stars and arrows for detections and upper limits respectively.
The nine unobserved post-mergers are shown with open stars and an
arbitrary atomic gas fraction log \fgas\ = $-2.0$.
For reference, galaxies from xGASS (Catinella et al. 2018) are also
shown -- grey filled symbols indicate xGASS detections and downward arrows are
upper limits.  The original GASS survey threshold is shown by the
 dotted line; the break in the \HI\ detection fraction threshold
at log (M$_{\star}/M_{\odot}$) = 10.5 was implemented to mitigate the large exposure
times required for a constant \fgas\ threshold at the lowest masses.  
However, it can be seen that at log M$_{\star}/ >$ 10.0 M$_{\odot}$
 the xGASS upper limits scatter around this threshold since the detectability
 of a source depends on its velocity width as well as its 21 cm flux.

For the purpose of this study, we define an effective detection threshold of
\fgas\ $ < $ 2 percent for log M$_{\star} >$ 10.5 M$_{\odot}$
and log \mhi = 8.8  M$_{\odot}$ below that mass.  The adapted detection
threshold is shown with a solid  line in Fig. \ref{fgas} and can
be seen to be a more conservative boundary between the detections and upper limits.
We stress that the exact placement of this definition does not qualitatively
affect the conclusions of this paper, it simply allows us to consistently
treat data points above and below the threshold.

From Fig. \ref{fgas} it is obvious by eye that the post-mergers occupy a region of preferentially
high \fgas\ at a given M$_{\star}$, and that there is an almost complete
absence of gas-poor post-mergers.  In the following sub-sections, we quantify the
difference between the post-mergers and xGASS in three different ways:
the \HI\ detection fraction, the median \fgas\ and the atomic gas fraction offset (from
a mass-matched control sample) and assess the
impact of including/excluding non-detections in these statistics.
We demonstrate that a full accounting of non-detections is critical
for a full assessment of gas fraction statistics and that a deep
\HI\ survey such as xGASS plays a crucial role in this measurement.

\subsection{HI detection fraction}

\begin{figure}
	\includegraphics[width=\columnwidth]{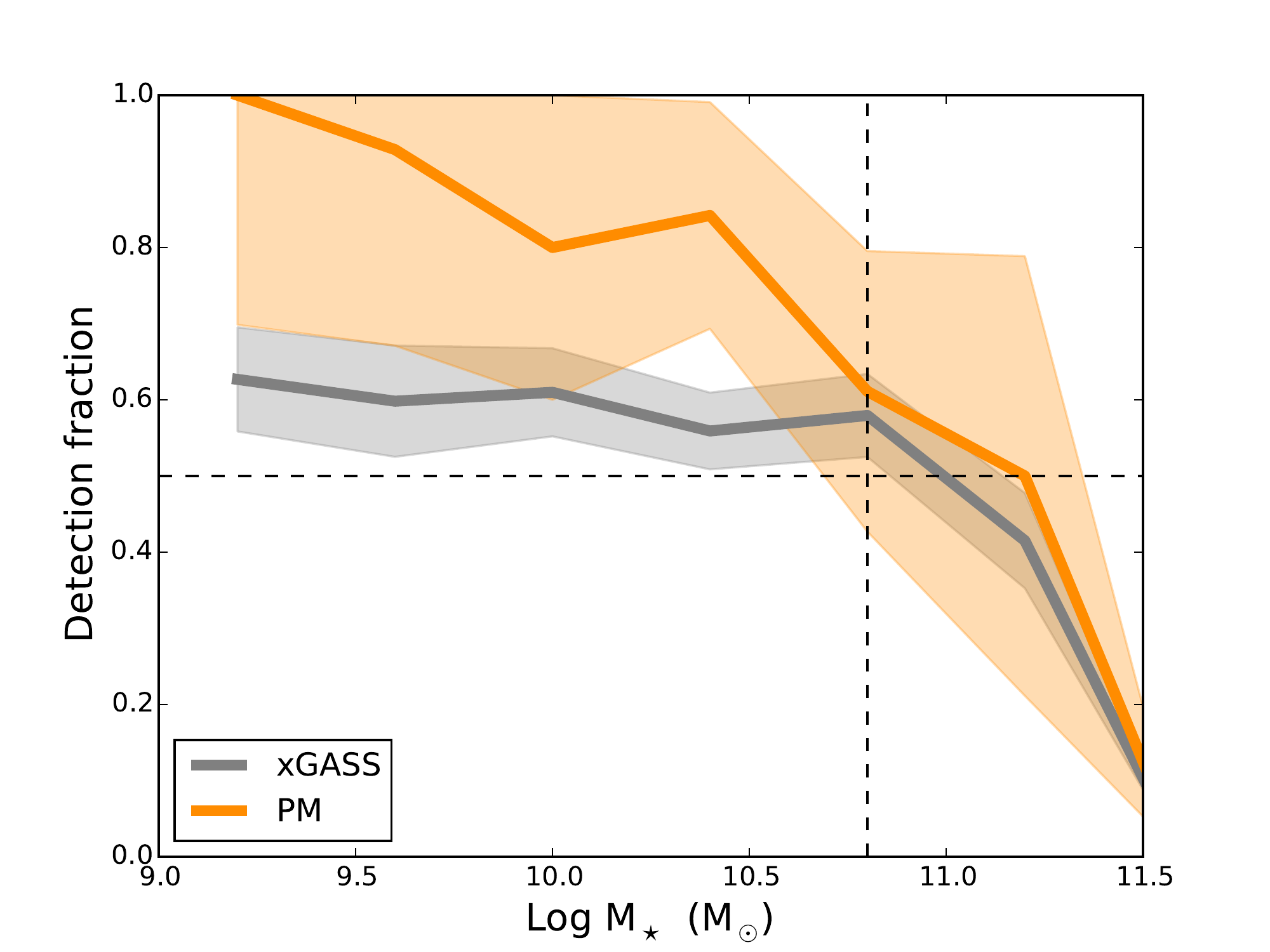}
        \caption{The \HI\ detection fraction of post-mergers (orange)
          and xGASS (grey) with 1$\sigma$ bounds shaded.  The vertical
          dashed line corresponds to log M$_{\star} =$ 10.8 M$_{\odot}$
          above which the xGASS detection fraction drops below 50 per cent
        (horizontal dashed line).}
    \label{det_frac}
\end{figure}

 In Fig. \ref{det_frac} we plot the \HI\ detection fraction
as a function of stellar mass for both the xGASS (grey line) and post-merger
(orange line) samples.  Galaxies are counted as non-detections if their upper
limit, or actual measured \fgas\ lies below the detection threshold shown
by the  solid line in Fig \ref{fgas}.  Upper limits above the detection
threshold (only 1 in our post-merger sample)
are treated as non-observations (i.e. not counted in the detection
fraction statistics).  The shaded regions show the $\pm$ 1$\sigma$ bound
and the horizontal dashed line shows a 50 per cent detection rate.  Both
the xGASS and post-merger samples show a steeply declining detection fraction
at high stellar masses.  However, the post-mergers have a systematically
higher 21 cm detection fraction at all stellar masses, a result that is most
significant at log M$_{\star} <$ 10.5 M$_{\odot}$, where the majority of the
post-mergers are located (Fig. \ref{fgas}) and hence the statistics are
most robust.  In this mass regime, the \HI\ detection fraction amongst
the post-merger sample is $\sim$ 50 per cent higher than the xGASS
control sample.  In a smaller, shallower sample of post-mergers
Ellison et al. (2015) also found a higher (factor of two) detection
fraction compared with their ALFALFA control sample.  

\subsection{Median f$_{\rm gas}$}

\begin{figure}
	\includegraphics[width=\columnwidth]{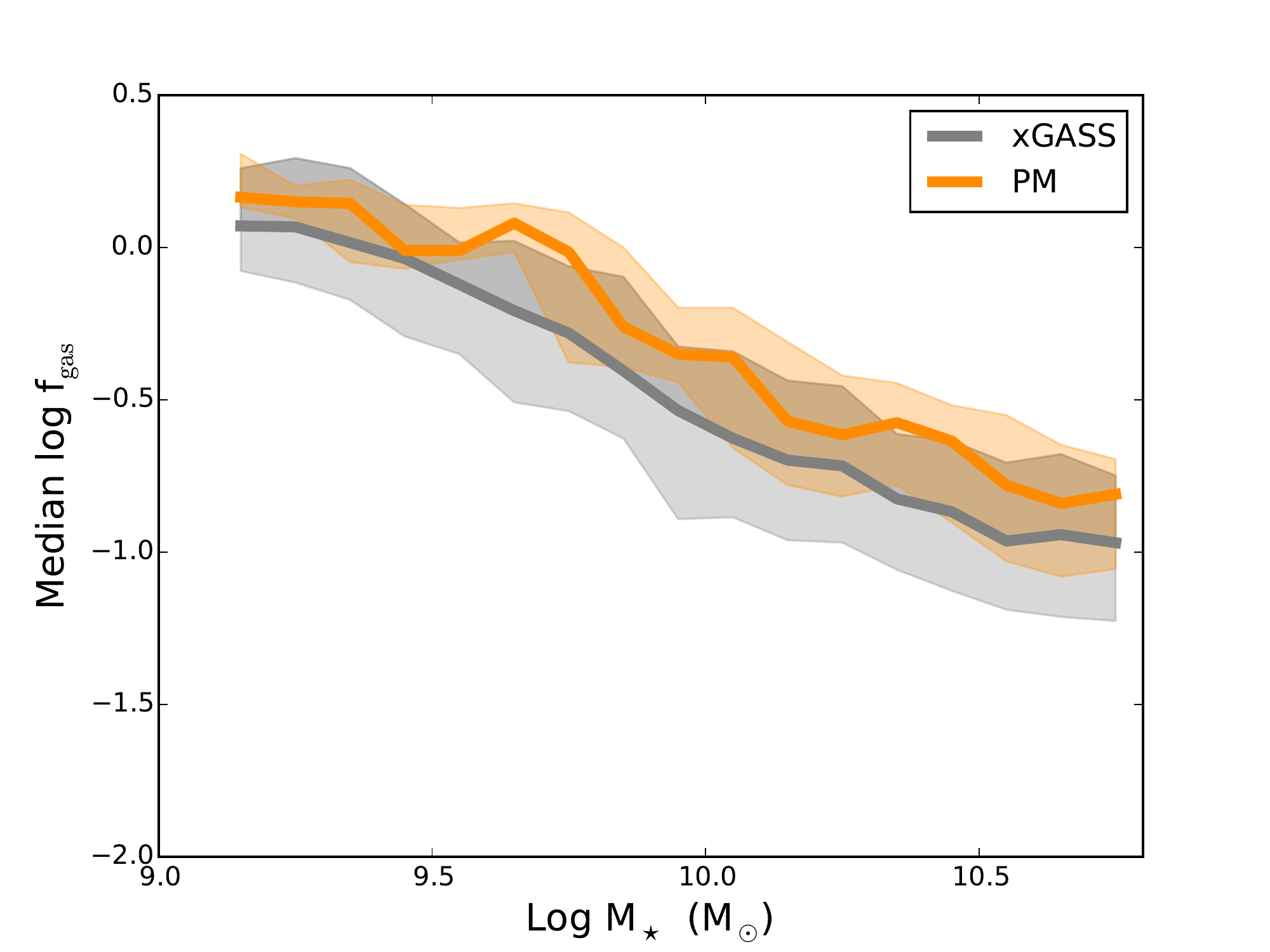}
	\includegraphics[width=\columnwidth]{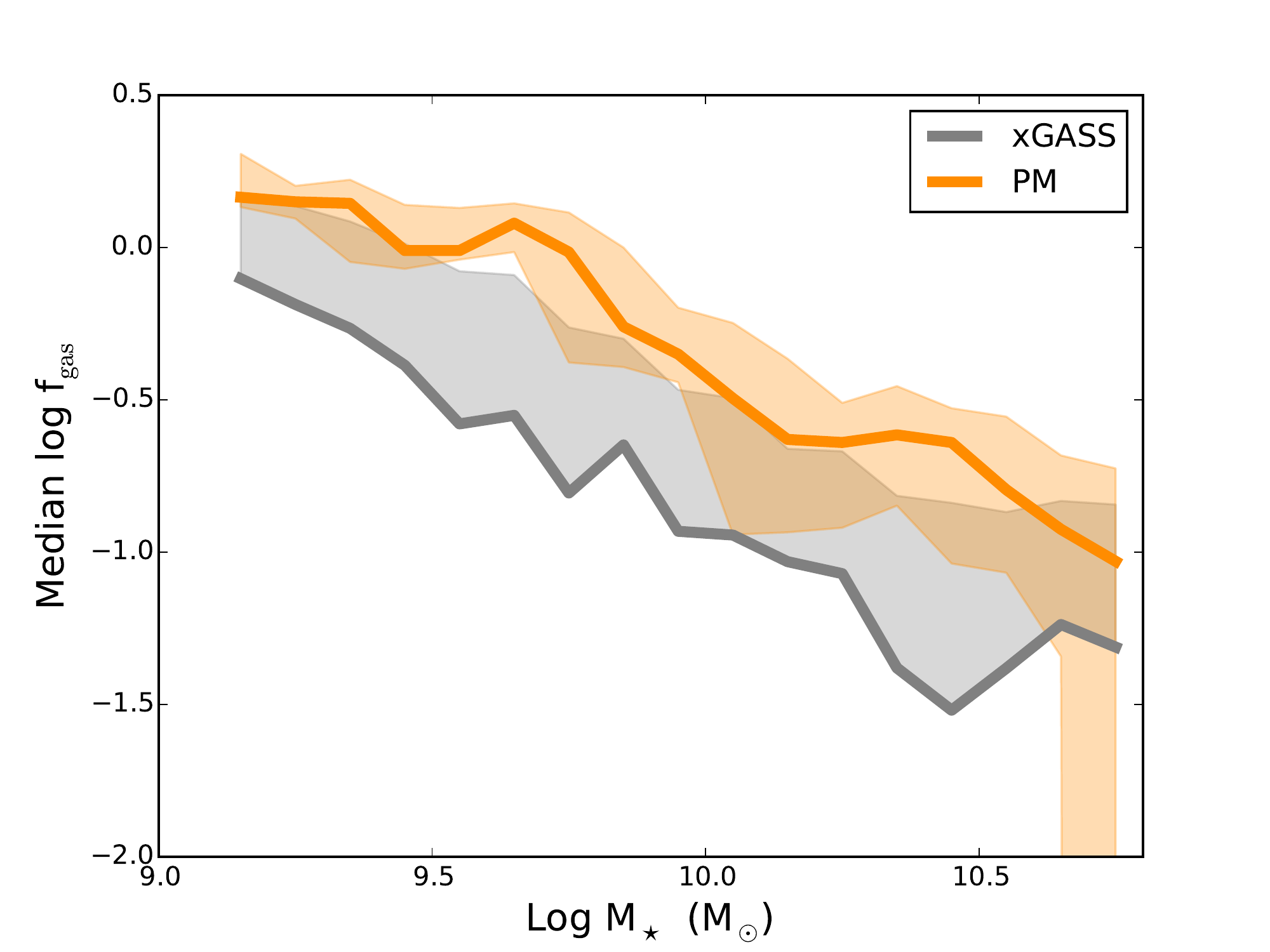}
        \caption{Median \HI\ gas fractions for post-mergers (orange)
          and xGASS (grey) with 25$^{th}$ and 75$^{th}$ percentiles shaded.
          Upper panel: The median \fgas\ is computed only including the
          \HI\ detections above the adopted detection threshold.
          Lower panel: The median \fgas\ is computed from both detections
          and upper limits. The 25$^{th}$ percentile for the xGASS sample
          in the lower panel can not be computed due to insufficient detections.}
    \label{fgas_med}
\end{figure}

In Fig. \ref{fgas_med} we compute the median \fgas\ for the xGASS and
post-merger samples, shown by the grey and orange lines, respectively.
The shaded zones in Fig. \ref{fgas_med} show the
25$^{th}$ and 75$^{th}$ percentiles of the \fgas\ distribution.
In the top panel of the figure we show the median \fgas\ for both samples only considering
galaxies for which there is an \HI\ detection.  In order to homogenize the
sample of detections, we only include detections that lie above our
adopted detection threshold (solid  line in Fig \ref{fgas}), ignoring
detections below this limit.  The median \fgas\ from \HI\ detections
in the post-merger sample
is systematically above the median value for the full xGASS sample, by
0.1 -- 0.2 dex, across the full range in stellar mass, although there
is considerable overlap between the 25$^{th}$ and 75$^{th}$ percentiles
of the two samples.

In Fig. \ref{det_frac} we showed that the \HI\ detection fraction of the
post-merger sample was significantly higher than xGASS.  Therefore,
the difference between the median \fgas\ of the two samples is likely
to be under-estimated if the non-detections are ignored.  In the
lower panel of Fig. \ref{fgas_med} we re-compute the median \fgas\
for the post-merger and xGASS samples now including the non-detections,
which yields a more complete and fair assessment of the difference
between the two samples.
Since Fig. \ref{det_frac} shows that the xGASS sample has a detection fraction $<$
50 per cent for log M$_{\star} >$ 10.8 M$_{\odot}$ (vertical dashed line
in Fig. \ref{det_frac}), it is only possible to determine a median \fgas\ for
the xGASS sample at stellar masses below this threshold.    Moreover, the
25$^{th}$ percentile can not be computed for the xGASS sample (at any stellar mass)
due to the insufficient detection fraction (as adopted in this work), hence the
lower grey shaded area is not plotted.  The lower panel of Fig.
\ref{fgas_med} again shows that the post-merger line lies systematically
above the xGASS sample.  The median \fgas\ of the post-merger sample
is little changed compared to the upper panel, because the \HI\ detection
fraction is high and therefore the median \fgas\ is not much changed
by the inclusion of non-detections.  However, the larger fraction of
non-detections in the xGASS sample leads to a significantly lower
median \fgas\ when the non-detections are included.  The difference
between the median \fgas\ in the post-mergers and xGASS
is now typically 0.3 -- 0.6 dex.

\subsection{Gas fraction offset: $\Delta f_{\rm gas}$}\label{dfgas_sec}

\begin{figure}
	\includegraphics[width=\columnwidth]{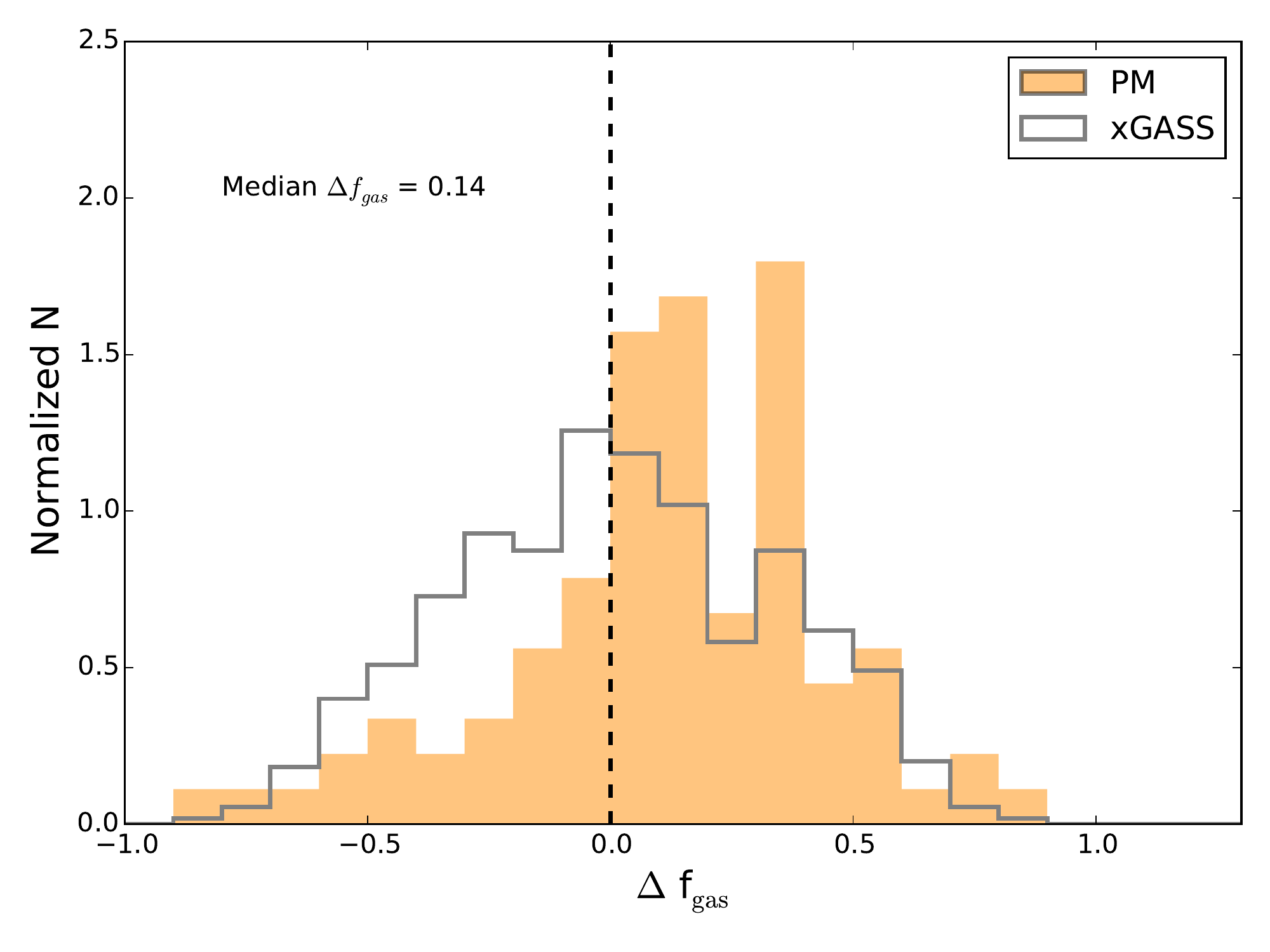}
	\includegraphics[width=\columnwidth]{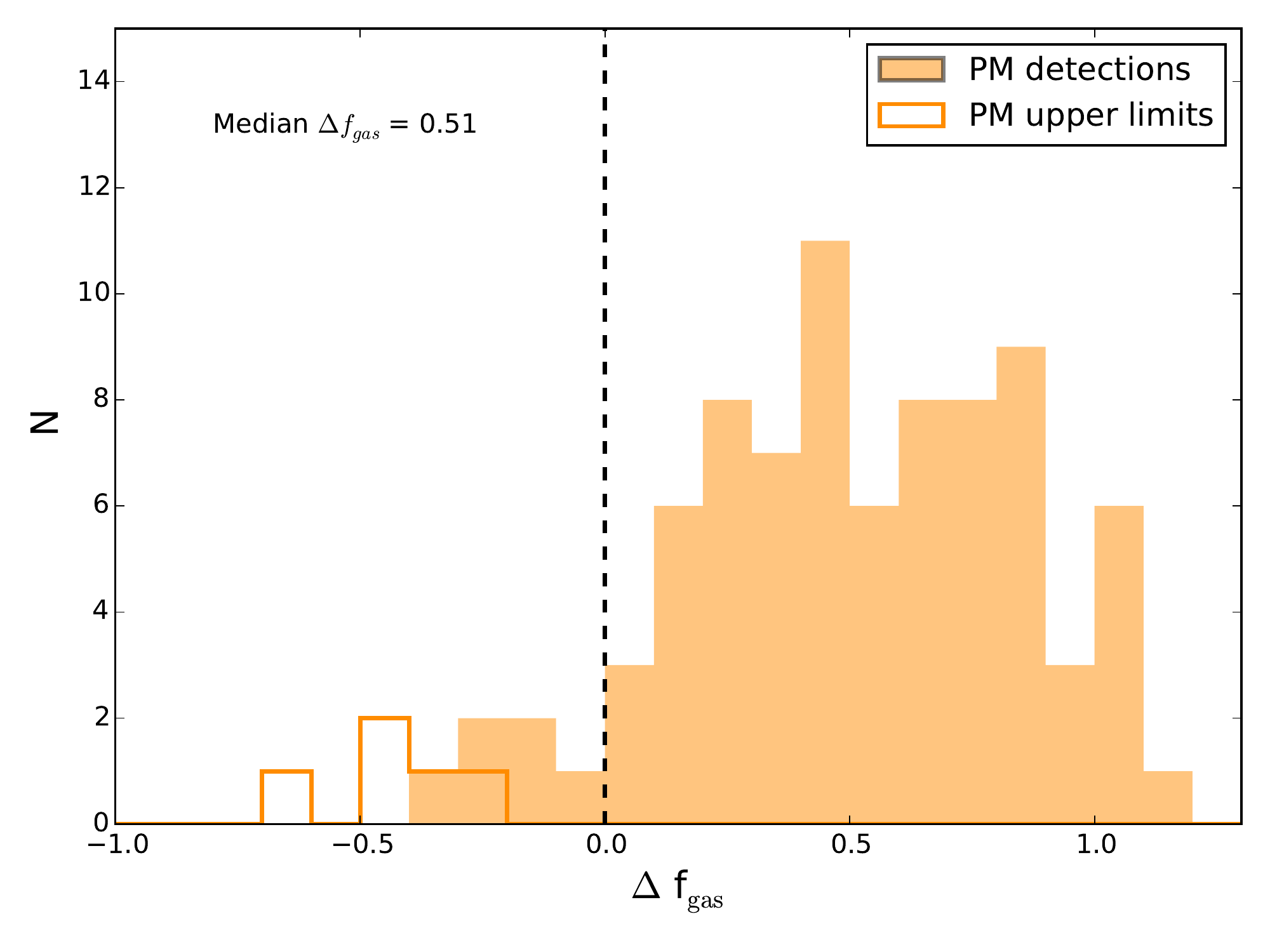}
        \caption{Atomic gas fraction offsets for the post-merger
          sample.  The median \dfgas\ for the post-mergers
          is given in each panel.  Upper panel: \dfgas\ distribution for the xGASS
          sample (open grey histogram) and post-mergers (solid
          orange histogram) based only on \HI\ detections above the
          adopted detection threshold.
          Lower panel:  \dfgas\ distribution for post-mergers
          with log M$_{\star} \le$ 10.8 M$_{\odot}$ considering both
          detections and upper limits in the control sample.
          The solid histogram shows \dfgas\ for \HI\ detections
          and the open histogram shows \dfgas\ upper
          limits for 5 non-detections in the same mass range.  Of
          the 9 unobserved post-mergers, 8 have masses log M$_{\star}
          \le$ 10.8 M$_{\odot}$ and have unconstrained \dfgas\
          (and are hence not plotted).  Note that the y-axis is normalized
          in the upper panel to facilitate a comparison with the (much
          larger) xGASS sample, but the unormalized histogram is shown
          in the lower panel to emphasize the majority of detections
        compared to non-detections and unobserved targets.}
    \label{dfgas}
\end{figure}

In order to quantify on a galaxy-by-galaxy basis the enhancement (or
deficit) of \fgas\ relative to the typical atomic gas fraction at that stellar
mass, we use the `gas fraction offset', \dfgas, first
introduced in our previous work on post-merger gas fractions (Ellison
et al. 2015). \dfgas\ is computed for each galaxy in turn by identifying
all galaxies in the xGASS sample of the same stellar mass (within $\pm$
0.15 dex) and computing the difference between log \fgas\ in the
post-merger galaxy and median value of the matched control galaxies.
Therefore, a galaxy with \dfgas\ = +1.0 has an atomic gas fraction 10 times
higher than expected for its stellar mass.  We require a minimum
of 5 mass-matched control galaxies per post-merger.  If this is not
achieved within the $\pm$ 0.15 dex tolerance, we grow the
tolerance by 0.1 dex and repeat until at least 5 matches are achieved.
In practice, typically $>$ 100 control galaxies are matched to each post-merger
without the need to grow the mass matching tolerance.

In Fig. \ref{dfgas} we show the distribution of \dfgas\ for the
post-mergers.  Following our analysis of the median \fgas,
we begin by showing in the upper panel the distribution
of \dfgas\ for the xGASS and post-merger sample considering
only \HI\ detections above the adopted detection threshold.
By definition, the distribution of \dfgas\ for xGASS is symmetric
around zero.  The post-mergers are biased towards positive
values of \dfgas; the median atomic gas fraction offset is \dfgas = +0.14,
indicating that, on average, the post-mergers are 40 per cent more
\HI\ rich than their mass-matched control sample.

However, we saw in the previous sub-section that considering
only detections for the xGASS control sample can under-estimate
the offset relative to the post-mergers, due to the higher fraction
of \HI\ non-detections in the former.  In the lower panel of
Fig. \ref{dfgas} we show the distribution of \dfgas\ for the
post-mergers that include non-detections from xGASS in the 
calculation of the control sample's gas fraction.    When including
\HI\ non-detections from the xGASS control sample, \dfgas\
can only be computed for galaxies with log M$_{\star} \le$ 10.8 M$_{\odot}$,
since the median \fgas\ of the xGASS control sample is not
constrained for higher masses (Fig \ref{det_frac}).
These \dfgas\ values are also reported in Table \ref{results_table}.

The solid histogram in the lower panel of Fig \ref{dfgas}
shows the \dfgas\ values for
the 83 post-mergers with log M$_{\star} \le $ 10.8 M$_{\odot}$
with \HI\ detections.  The 5 non-detections in this stellar mass
range, which correspond to upper limits in \dfgas, are shown
by the open histogram.  Of the 9 unobserved targets in the final
post-merger sample, 8 have log M$_{\star} \le $ 10.8 M$_{\odot}$.
These are not plotted in Fig. \ref{dfgas} and could be located
at any value of \dfgas. However, the overwhelming majority
of the post-merger sample have positive values of \dfgas,
such that the 8 unobserved targets will not influence the conclusion
that the post-mergers are gas-rich.  Once the non-detections
have been included in the control sample, the median gas
fraction offset is \dfgas\ = +0.51 dex, indicating that the post-mergers
are typically 3 times more abundant in atomic gas than the
control sample.

\section{Discussion}\label{discussion_sec}

\subsection{Comparison with previous results}

In our previous study of the \HI\ content of post-mergers,
Ellison et al. (2015) found that the gas fractions in a
sample of 37 post-mergers were consistent with a control sample
drawn from the ALFALFA survey.  However, we
also found that the post-merger sample exhibited a factor of two
higher \HI\ detection fraction than the ALFALFA control sample.
The combination of these results led us to
speculate that the shallow survey depth of the dataset,
enforced by the adoption of ALFALFA as the comparison sample,
may have led to an under-estimate of the atomic gas fraction offset
in post-mergers.  Ellison et al. (2015) concluded that, contrary
to being gas depleted after an interaction-induced starburst,
post-merger galaxies may actually be relatively gas-rich.

In the new work presented here, we have confirmed the
speculation of Ellison et al. (2015), thanks to a larger,
deeper \HI\ survey of post-mergers and the availability
of an equivalently deep comparison sample: xGASS. In Figures
\ref{fgas_med} and \ref{dfgas} we have demonstrated
the importance of including \HI\ non-detections and shown
that the median \fgas\ and \dfgas\ are under-estimated
if only detections are used.  Once non-detections are accounted
for, post-merger galaxies have, on average, a higher \HI\ gas fraction by
a factor of 3, compared with a mass-matched control
sample in xGASS (Catinella et al. 2018).

Our work is consistent
with that of Janowiecki et al. (2017) who found that centrals
in groups of two (i.e. pairs) in xGASS had a factor of two
higher \fgas\ than isolated central galaxies of the same stellar mass. Other
surveys that have reported elevated \HI\ gas fractions in mergers include
Casasola et al. (2004), Huchtmeier et al. (2008) and Jaskot et al.
(2015).  Conversely, based on a sample of dwarf galaxy pairs,
Stierwalt et al. (2015) found \HI\ gas fractions consistent
with their control sample.  However, since the Stierwalt et al.
(2015) control sample was drawn from ALFALFA, the comparison
may suffer from the same limitation as that of Ellison et al.
(2015).  The new results presented here highlight the importance
of sensitive \HI\ measurements and the need to take into
account non-detections.

\subsection{The connection with SFR and asymmetry}\label{sfr_sec}

Analogous to the enhanced atomic gas fractions in merging galaxies
presented here, elevated fractions of \textit{molecular} gas have also been
reported in interacting galaxies of various stages (Braine \& Combes
1993; Combes et al. 1994; Casasola et al. 2004).  The
most readily comparable study to our investigation of atomic gas are
the studies of molecular gas in close galaxy pairs
by Violino et al. (2018), Pan et al. (in prep) and in post-mergers by
Sargent et al. (in prep.).
These works have also drawn their merger samples from the SDSS
and used comparable matching techniques to the ones presented
here in order to compute gas fraction offsets in the molecular phase.
Violino et al. (2018) and Pan et al. (in prep) 
measure a molecular gas fraction \dfgash2\ = 0.4 dex
for close galaxy pairs and Sargent et al. (in prep) find a 0.6 dex
enhancement for the post-mergers, compared with a stellar mass matched
sample.   Both Violino et al. and Sargent et al. find that the molecular
gas fraction enhancement is reduced if SFR is used as an additional
matching parameter (although \dfgash2 = 0.2 dex for the post-mergers even
after this additional matching).  Indeed, it is well known that
(at fixed M$_{\star}$) galaxies with higher SFR have higher H$_2$
fractions (Saintonge et al. 2012, 2016).  

A galaxy's atomic gas fraction has also been shown to correlate
with its SFR.  For example, Brown et al. (2015) show that, at
fixed stellar mass, galaxies with bluer $NUV-r$ colours (a proxy
for specific SFR) have higher gas fractions (see also Saintonge et al. 2016).
We therefore repeated our \dfgas\ calculations for the post-mergers, matching
in both stellar mass and total SFR, the latter of which is also taken
from the MPA/JHU catalogs (e.g. Brinchmann et al.
2004; Salim et al. 2007).  Although there may be concern that
significant star formation may be obscured in merging galaxies,
Sargent et al. (in prep) have performed
a thorough testing of the SFRs in a smaller sample of post-mergers
selected from SDSS, and find consistent results with SFRs derived
from the UV+IR.  In addition to the baseline
mass matching tolerance of 0.15 dex, we also now require
that the total SFR of the control galaxies matches the post-merger
within 0.1 dex.  If less than 5 control galaxies are simultaneously
matched in mass and SFR, both quantities have their matching
tolerances increased by 0.1 dex.  Most post-mergers have at
least 5 control galaxies without the need to expand the matching
tolerances.  However, whilst the typical number of controls per
post-merger was $>$ 100 when matching in mass alone, this number
decreases to 10--20 for simultaneous matching with SFR.

\begin{figure}
	\includegraphics[width=\columnwidth]{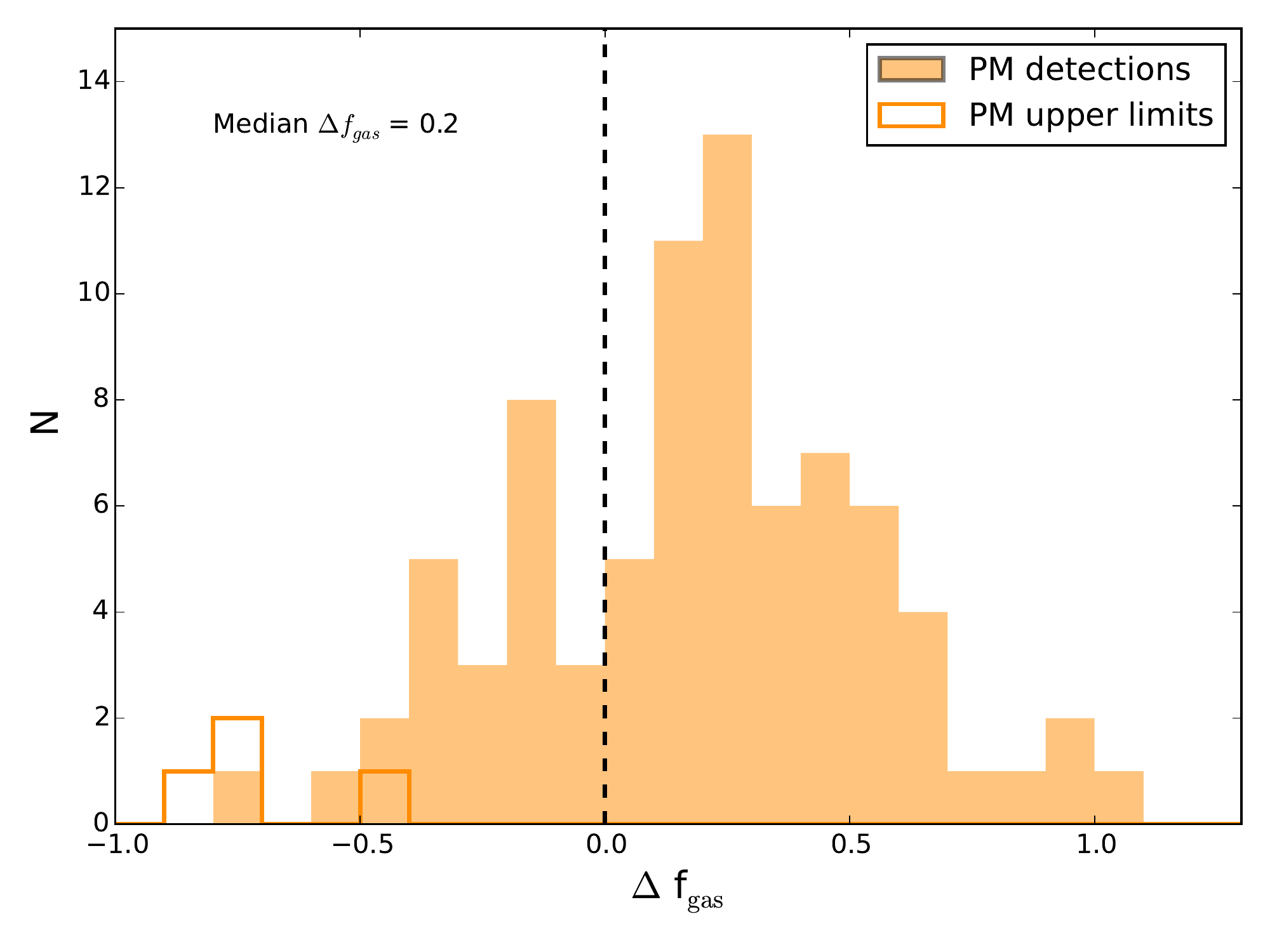}
        \caption{  \dfgas\ distribution for post-mergers
          with log M$_{\star} \le$ 10.8 M$_{\odot}$ with the control sample
          matched in both stellar mass and SFR.  
          The solid histogram shows \dfgas\ for \HI\ detections
          and the open histogram shows \dfgas\ upper
          limits for the non-detections in the same mass range.  }
    \label{dfgas_v5}
\end{figure}

In Fig. \ref{dfgas_v5} we show the distribution of \dfgas\ for
the mass and SFR-matched control sample.  Fig. \ref{dfgas_v5}
includes \HI\ non-detections in both the post-merger and control sample
and is hence comparable to the lower panel of Fig \ref{dfgas}.
The additional match in SFR has reduced the median \dfgas\
from +0.51 dex (lower panel of Fig. \ref{dfgas}) to +0.2 dex (Fig \ref{dfgas_v5}), which
closely mirrors the molecular gas fraction offsets in post-mergers
with and without SFR matching found by Sargent et al. (in prep.).
Fig. \ref{dfgas_v5} shows that even accounting for the elevated
SFRs in mergers, an excess of \HI\ gas, relative to the control sample,
remains.  

\begin{figure}
	\includegraphics[width=\columnwidth]{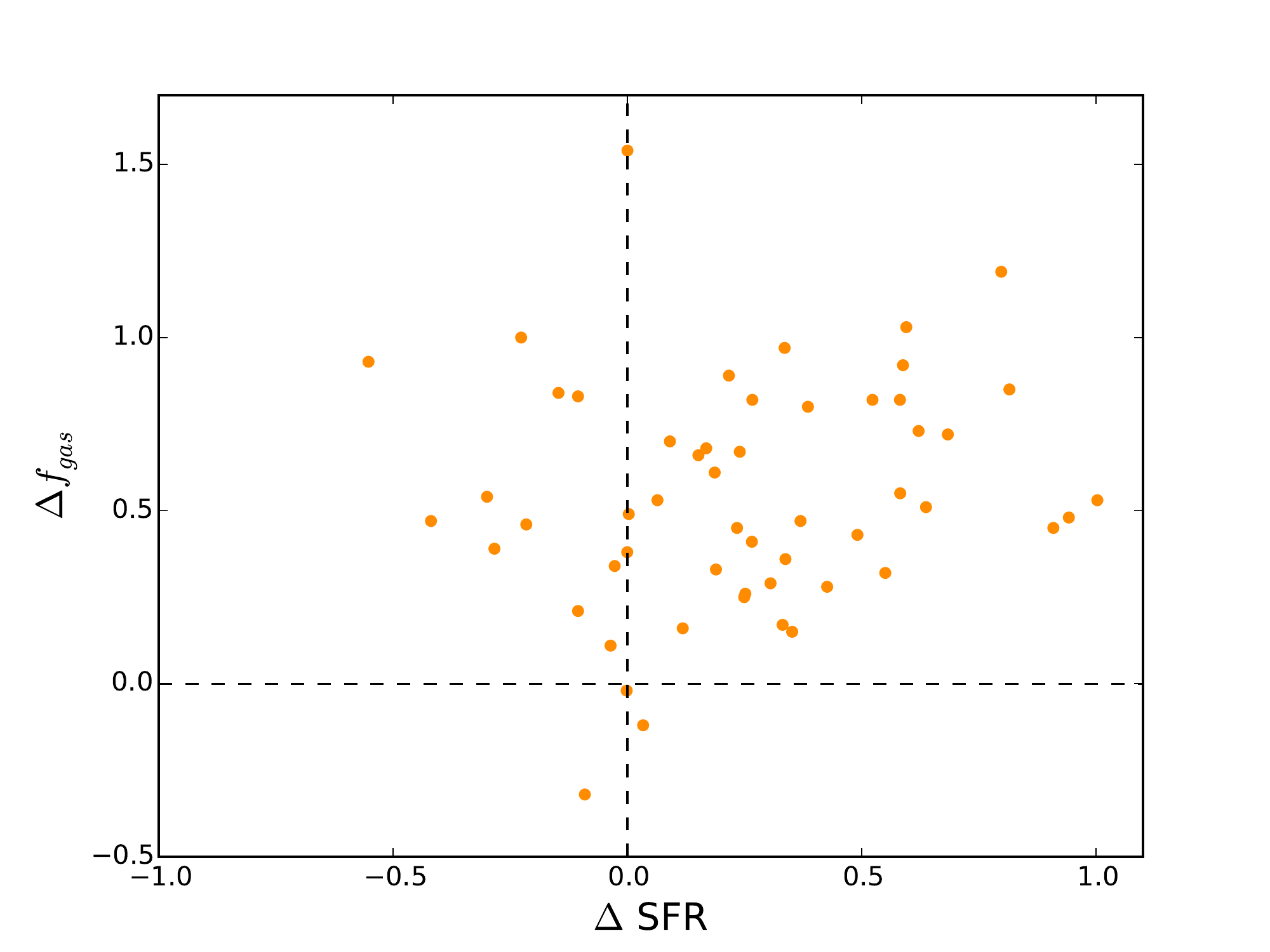}
        \caption{SFR offsets relative to the main sequence ($\Delta$ SFR)
versus \dfgas.  Whilst the post-mergers in our sample have generally
elevated SFRs, there is no correlation between \dfgas\ and $\Delta$
SFR.  Post-merger galaxies exhibit elevated \HI\ gas fractions over
a wide range of relative SFRs.}
    \label{dsfr}
\end{figure}

In Fig. \ref{dsfr} we further investigate the connection between SFR and
enhanced atomic gas fractions by plotting \dfgas\ versus $\Delta$ SFR, which is defined as
the SFR relative to main sequence galaxies of the same
stellar mass, redshift and local environment (see Ellison
et al. 2018 for more details on the calculation of $\Delta$ SFR).
Fig. \ref{dsfr} shows that, as known from Ellison et al. (2013), the
post-merger sample has generally elevated SFRs (i.e. positive $\Delta$
SFR).  However, Fig. \ref{dsfr} also shows that
there is no correlation between $\Delta$ SFR and \dfgas\  (the
Pearson correlation coefficient is 0.15); post-merger
galaxies show enhanced \HI\ gas fractions over a wide range of
SFR offsets.

We also checked for a possible correlation between galaxy
asymmetry and \dfgas.  Galaxy asymmetry may be used as
a rough indicator of the time since the merger, although
it is also sensitive to the physical properties of the
progenitors and their orbits (Lotz et al. 2010a,b; Ji, Peirani \& Yi 2014).
We took the $r$-band asymmetry index from Simard et al. (2011),
which measures the fraction of the light in the asymmetric
component after the subtraction of a bulge+disk model.
We found no relationship between asymmetry and \dfgas, confirming
a similar result in our earlier work (Ellison et al. 2015).

\subsection{The connection with AGN}

The \HI\ CGM content of galaxies has recently been found to be
enhanced in galaxies that host AGN (Berg et al. 2018).
Based on a sample of 20 AGN host galaxies selected from the
SDSS, Berg et al. (2018) measured enhanced \lya\ absorption in the
CGM at impact parameters  $\sim$ 160 -- 300 kpc.  Berg et al. (2018)
speculate that galaxies that host AGN may have preferentially
higher gas fractions.  There is a well documented connection between
galaxy mergers and AGN triggering (Ellison et al. 2011; Satyapal
et al. 2014; Khabiboulline et al. 2014), which could indicate
that mergers are responsible for the enhanced CGM \HI\
content observed by Berg et al. (2018).  Of the post-mergers in
our sample, 25 are classified as optical AGN (Kauffmann et al.
2003). However, the
\dfgas\ distribution of these 25 optically selected AGN in
our post-merger sample is indistinguishable from the
atomic gas fraction offsets of rest of the sample (see also
Fabello et al. 2011).  This does
not necessarily rule out a connection between merger-triggered
AGN and elevated atomic gas fractions, since AGN life-times can
be short and flicker on the timescales associated with
the post-merger stage (e.g. Schawinski et al. 2015).

\subsection{Environmental dependence}

It is well known that the \HI\ content of galaxies depends
on large scale environment, with lower gas fractions
in both cluster and group environments (Solanes et al.
2001; Chung et al. 2009; Kilborn et al. 2009; Cortese et al. 2011;
Hess \& Wilcotts 2013).  Galaxies classified as satellites
within their halo (Yang et al. 2007) appear to be the most
\HI\ deficient at a given stellar mass (Catinella et al. 2013).
Of the 107 post-mergers in our sample, 57 are classified as 
isolated centrals, 23 are group centrals, 23 are satellites 
and 4 are not in the Yang et al. (2007) catalog.  The post-merger
satellite fraction (22 per cent) is therefore somewhat lower
than the xGASS control sample in general (32 per cent), indicating
that environmental differences might lead to lower atomic gas fractions
in the control.

In order to test for trends with halo class, we repeated 
our calculation of \dfgas, separating both the control and
post-merger sample into 3 categories based on Yang et al. (2007):
satellite, group central and isolated central.  Only controls
of the same halo type are matched to each post-merger, i.e.
a post-merger galaxy that is classified as a satellite
only has controls that are also classified as satellites.  We
additionally attempted to match in halo mass, but the statistics
for this proved insufficient.    We restrict our calculation of \dfgas\ to the 83 galaxies
with log M$_{\star} < 10.8$ and \HI\ detections (i.e. the solid histogram
in Fig \ref{dfgas}), ignoring the 5 non-detections in this stellar mass
range.

\begin{figure}
	\includegraphics[width=\columnwidth]{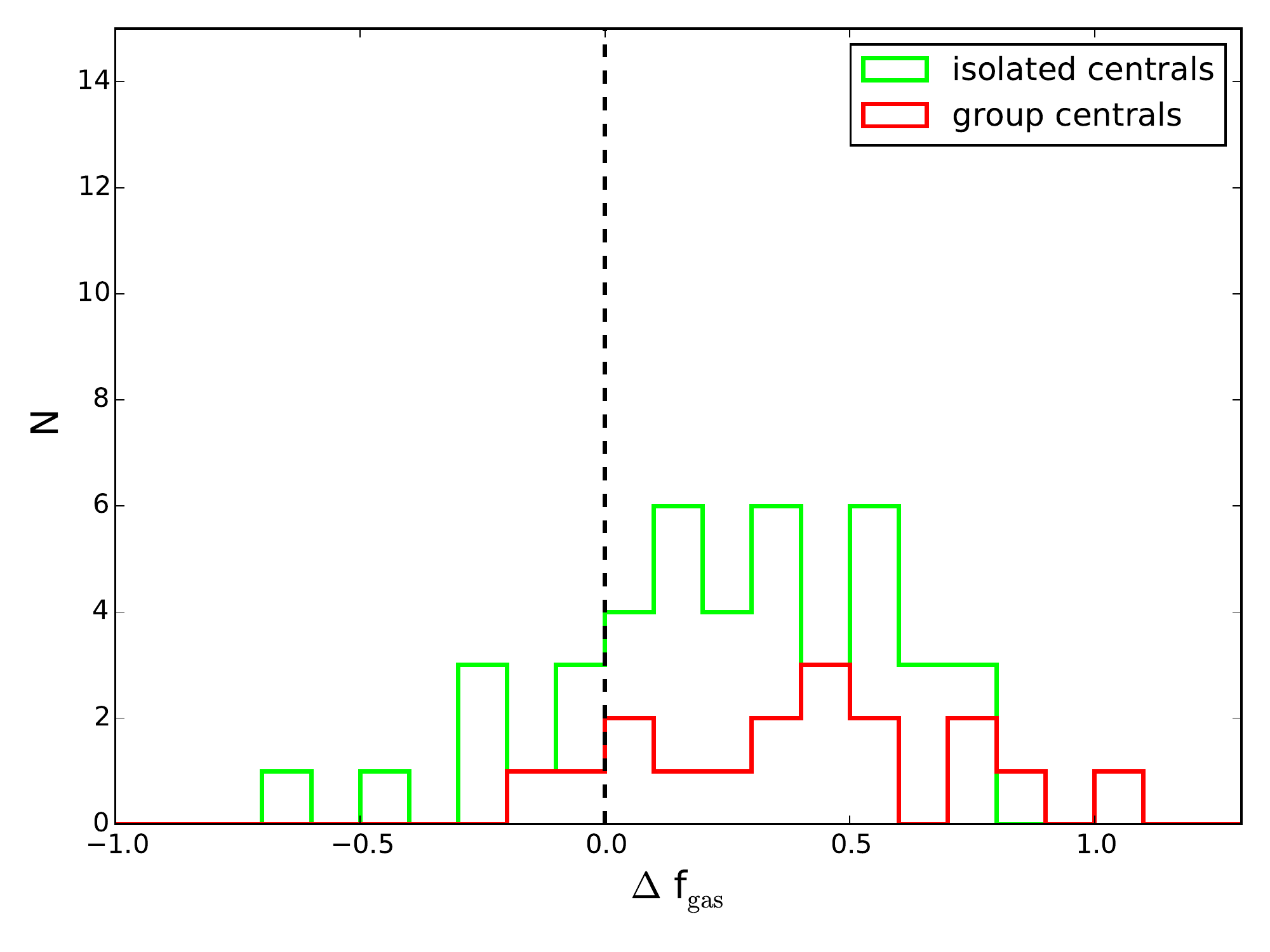}
        \caption{Atomic gas fraction offsets for the post-merger
          sample as a function of halo type (taken from
Yang et al. 2007). Atomic gas fraction offsets are computed including
non-detections in the xGASS control sample.  Values of \dfgas\ for
post-mergers classified as satellites are not constrained
due to the low \HI\ detection fraction in the control sample.  }
    \label{dfgas_halo}
\end{figure}

Fig. \ref{dfgas_halo} shows the
\dfgas\ distribution for group and isolated centrals 
in the post-merger sample, both of which are still strongly
skewed towards positive values, despite the additional
environmental matching.  Unfortunately, \dfgas\ is unconstrained for all of the post-merger 
satellite galaxies due to a $>$ 50 per cent non-detection rate
amongst their controls.  We have therefore not found
any evidence that the high \HI\ gas fractions in the post-merger 
sample are the result of an environmental bias on the halo scale.
However, our environment testing is limited by available data, and
we have not been able to investigate halo mass effects, nor the impact
on satellite galaxies, for the reasons described above.  Furthermore,
 larger scale environmental effects, such as inflowing
gas from the IGM are not accounted for in these tests.

\subsection{Why are post-mergers HI rich?}

In Ellison et al. (2015) we discussed possible reasons
behind the enhanced atomic gas fraction in post-merger galaxies, and
re-visit that discussion here in light of both new data and
simulations.  First, we re-consider the suggestion put forward
by Ellison et al. (2015) that post-mergers might naturally
be \HI\ enhanced due to the arithmetic combination of the
progenitor sample.  We repeat the experiment of Ellison et al.
(2015) in which we produced a fake sample of post-mergers
from their control sample.  Whereas Ellison et al. (2015)
did this experiment using the shallow (and hence, gas-rich)
ALFALFA sample, we will do it using the xGASS sample.
We draw at random two sets of 83 galaxies (the number of
\HI\ detections in our post-merger sample) from the xGASS
detections and sum their stellar and atomic gas masses, to emulate
the arithmetic combination occuring in a merger.  We then
compute the \dfgas\ of the resulting post-merger sample.
This process is repeated 1000 times with the median \dfgas\
of each fake post-merger survey recorded.

\begin{figure}
	\includegraphics[width=\columnwidth]{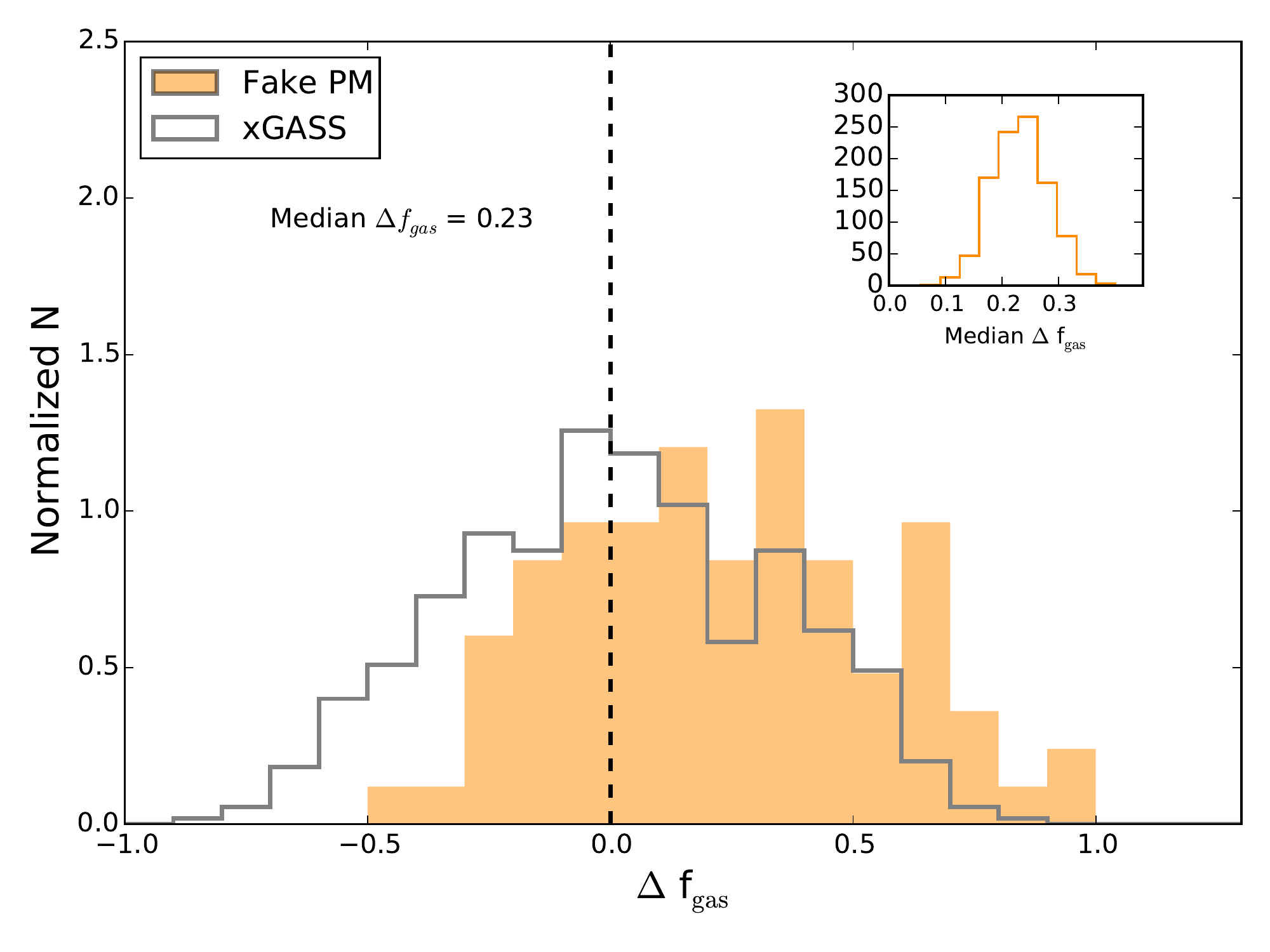}
        \caption{Results of simulating 1000 fake post-merger
surveys each with 83 \HI\ measurements.  The filled orange 
histogram shows the distribution
of \dfgas\ that results from one fake survey generation in which
83 pairs of randomly selected xGASS galaxies with \HI\ detections are combined in
stellar and \HI\ mass.  The original \dfgas\ distribution for
the xGASS sample is shown in the open grey histogram (replicated
from Fig. \ref{dfgas}).  The inset histogram shows the 
(unnormalized) distribution
of median \dfgas\ from the 1000 survey generations.  The typical
expected median \dfgas\ simply from combining pairs of xGASS galaxies
is $\sim$ 0.25 dex.}
    \label{sim_dfgas}
\end{figure}

In the main panel Fig. \ref{sim_dfgas} we show the distribution of
one fake post-merger survey generation, consisting of
83 galaxies as the filled orange histogram.
For reference, the initial xGASS distribution
of \dfgas\ is shown in grey; this is the same distribution shown
in the top panel of Fig. \ref{dfgas} and is symmetric around zero
by construction.  The median \dfgas\
of this iteration is 0.23 dex.  The inset
histogram shows the distribution of median \dfgas\
from the 1000 fake survey generations.  The typical
expected median \dfgas\ simply from combining pairs of xGASS 
galaxies is $\sim$ 0.25 dex.  This result is qualitatively consistent
with, although slightly larger than, 
the observed atomic gas fraction enhancement measured in
the real post-merger sample when only considering detections
(\dfgas\ = 0.14 dex, Fig. \ref{dfgas}).  This experiment indicates that we
might naturally expect post-mergers to have higher atomic gas
fractions than their mass matched control sample. 

Even if an arithmetic combination of galaxy pairs is
expected to yield a positive atomic gas fraction offset, this
enhancement might be mitigated by the elevated SFRs
(and subsequent depletion/expulsion of the gas reservoir)
experienced by galaxy mergers.  In Ellison et al. (2015)
we made a crude assessment of this by using a suite of
simulated binary mergers and assuming that the enhanced
star formation could be used to estimate the maximum
depletion of the \HI\ reservoir.  However, since the simulations
used by Ellison et al. (taken from Moreno et al. 2015) did
not account for the different ISM gas phases, we could not
distinguish different gas phases and trace the transition
between them.

Moreno et al. (2018) have recently performed a new suite of
binary merger simulations using the state-of-the-art
FIRE-2 code, which aims
to recover the detailed physics of the ISM at high resolution (Hopkins
et al. 2018).  Moreno
et al. (2018) study a range of galaxy merger orbits and
trace the multi-phase ISM during the pair (pre-coalescence) phase 
of 27 interactions.  The molecular gas reservoir is found to be
enhanced throughout the interaction phase, by 0.1 dex on average, which in turn fuels
additional star formation.  This is qualitatively consistent with observational
studies of enhanced molecular gas fractions in pairs (Violino et al. 
2018). Early in the simulated interactions, the enhanced molecular gas fraction
is mirrored by a decline in the atomic gas fraction, as the latter's
reservoir fuels the former.  However, the atomic gas fraction
is steadily replenished throughout the simulated interactions 
as hot gas cools.  Therefore, despite the enhanced star formation
rates that are observed in simulations and observations alike,
the results from Moreno et al. (2018) indicate that we might
not expect to see this draining the atomic gas reservoir. Instead,
the results from Moreno et al. (2018) indicate that if we
see any change in \HI\ it could be a small enhancement - the median
change in HI gas fraction in their suite is a 5 per cent increase.
Further \HI\ replenishment could come from cooling of a hot halo
(e.g. Moster et al. 2011).
We therefore propose that the observed enhancement in \fgas\
is dominated by the arithmetic combination of the progenitor's
atomic gas reservoirs, with a possible minor contribution from 
the net enhancement of the atomic component from cooled ionized gas.

Finally, we consider whether the high atomic gas fractions in our
post-merger sample could be due to a selection bias.  The post-merger
sample is selected based on visual classifications of asymmetry
and disruption.  Simulations have shown that such strong tidal
features are more prevalent and longer lived in interacting galaxies
with high gas fractions (Lotz et al. 2010a).  It is therefore possible
that the progenitors of our post-merger sample had pre-existing
high gas fractions.  Indeed, the images of our post-mergers
(Figs \ref{detect_fig} and \ref{app_det_fig}) indicate their progenitors were
more likely to be disks than ellipticals.  We have therefore
repeated the \dfgas\ analysis presented in Sec. \ref{dfgas_sec} using only
control galaxies that are disk dominated.  This is achieved by imposing an
$r$-band bulge fraction cut of $<$ 0.3 (Simard et al. 2011) onto the
control pool of xGASS galaxies.  We find our results are largely
unchanged, with the resulting median \dfgas\ still at $\sim$ 0.5 dex.
Therefore, whilst we can not rule out that the progenitor galaxies of
our post-merger sample might have been already gas-rich, we have not
found evidence of this bias.

Our results have shown that elevated \HI\ gas fractions can be achieved
through the merger process, which raises the question of whether
\textit{all} high atomic gas fraction galaxies are post-mergers.  To investigate
this, we computed \dfgas\ values for xGASS galaxies and visually
inspected the SDSS images of the upper quartile of the sample,
i.e. those with the largest atomic gas fraction enhancements.  These
galaxies rarely show any signs of interactions, indicating that
elevated \fgas\ can also be achieved through non-merger processes
(see also Sancisi et al 2008; Gereb et al. 2018). Alternatively,
the high \fgas\ galaxies in xGASS \textit{may} have been due
to a past merger, but their optical asymmetries have now faded.
Indeed, Lutz et al. (2018) have recently found that warps and \HI\
asymmetries are more common in gas-rich galaxies.

\subsection{Do galaxy mergers lead to quenching?}

\begin{figure*}
\includegraphics[width=18cm]{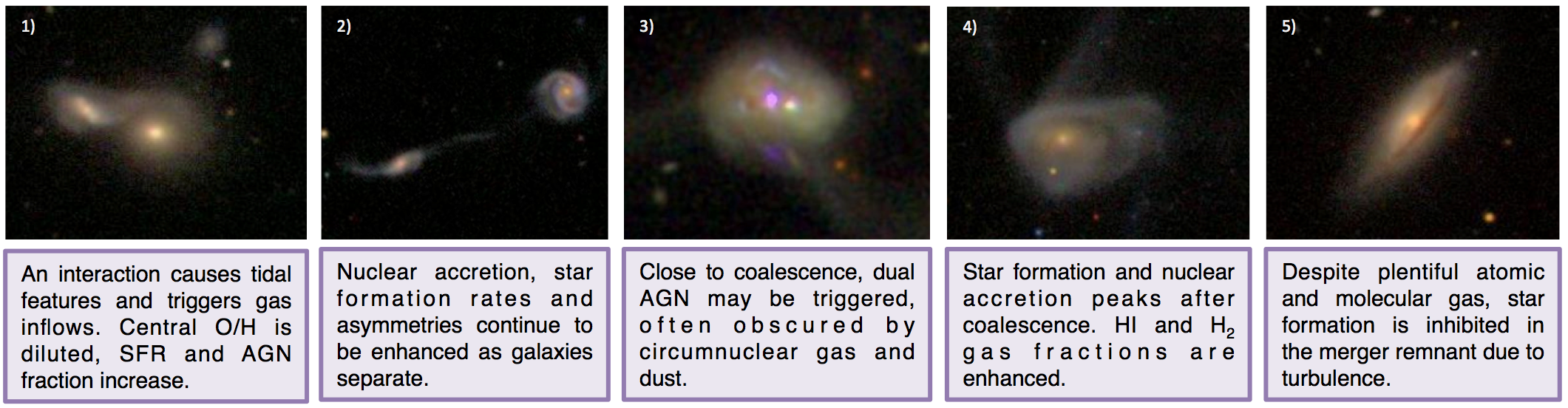}
\caption{A revised picture of the galaxy merger process based on observations from the
SDSS and complementary multi-wavelength data.  Importantly, we find no evidence for
gas 'blowout' and propose that if star formation declines following a merger, it is
more likely to be due to the inability of the extant gas reservoir to form stars,
due to increased turbulence.}
\label{evo_fig}
\end{figure*}

Simulations have played a pivotal role in our understanding of galaxy mergers, 
having successfully predicted many facets of the interaction
process.  For example, merger simulations predict (and can model) morphological
asymmetries (e.g. Pop et al., 2018; Mortazavi et al. 2016), boosts in SFR (e.g.
Di Matteo et al. 2007; Cox et al. 2008; Moreno et al. 2015),
enhanced nuclear accretion (Capelo et al. 2015; Blecha et al. 2018) and
changes in metallicity gradients (Perez et al. 2011; Torrey et al. 2012;
Bustamante et al 2018).  Importantly, simulations have also predicted
that a wholesale removal of gas can occur, either due to ejection
by powerful winds (e.g. Di Matteo et al. 2005; Hopkins et al. 2008) or consumption by
star formation (e.g. Bekki 1998).  This potentially key phase
in the merger process, which links mergers to the quenching 
of star formation, remains unproven by observations.

The results presented here demonstrate that, contrary
to being gas depleted, recently merged galaxies exhibit an excess
of atomic hydrogen gas.  A similar excess (at fixed stellar mass)
is also present in the molecular phase, both before and
after the interaction is complete (Braine \& Combes 1993;
Casasola et al. 2004; Violino et al. 2018; Pan et al. in prep; Sargent et al. in prep).
\textit{Taken together, these observational studies indicate that
the merger process is unlikely to lead to the quenching of star formation
as a result of gas exhaustion or expulsion.}  Moreover, our results
indicate that the enhanced molecular gas fractions do not simply result
in a net depletion of the atomic gas reservoirs.  
Indeed, Jaskot et al. (2015) find
that starburst galaxies have very similar \HI\ gas fractions to
`normal' star forming galaxies.
The simulations of Moreno et al. (2018) show that these observations
can be understood via the complex conversion of gas between
different phases.  Specifically, although atomic gas does feed
the molecular phase during the interaction, the atomic
component can also be replenished by cooling from the ionized phase.

One caveat to our conclusion that mergers do not lead to atomic gas
depletion is that the post-mergers in our
sample have been selected based on strong morphological asymmetries,
and hence likely represent fairly recently coalesced systems.
Their SFRs are still enhanced (e.g. Fig. \ref{dsfr}) and studies
of similarly selected systems have shown nuclear accretion is
on-going (e.g. Satyapal et al. 2014; Weston et al. 2017).  It is
therefore possible that gas blowout/consumption is yet to happen.
However, there are several complementary results that support our
conclusions.  First, Fabello et al. (2011) have shown that the
\HI\ gas fractions of AGN are consistent with non-AGN host galaxies,
and conclude that the AGN therefore has little impact on its
galactic gas reservoir.  Indeed, Berg et al. (2018) find slightly
\textit{higher} Ly$\alpha$ absorption equivalent widths in quasar
sightlines that probe local Seyfert hosts.  Regarding gas consumption, 
several studies have shown that post-starburst galaxies retain a 
considerable molecular (French et al. 2015; Rowlands
et al. 2015; Alatalo et al. 2016; Suess et al. 2017; Smercina et al.
2018) and atomic (Buyle et al. 2006; Zwaan et al. 2013) gas reservoirs.
Since a significant fraction of post-starbursts seem to 
be related to late stage mergers (Zabludoff et al. 1996; Goto
2005; Meusinger et al. 2017; Pawlik et al. 2018) their substantial gas
reservoirs indicate that the gas remains even after an
interaction-induced starburst is complete.
Finally, since 40 per cent of early-type galaxies have been shown to
maintain significant molecular  gas reservoirs (e.g. Young et al. 2014),
it seems that the cessation of star formation and morphological
transformation do not always go hand-in-hand with an absence of gas.

If the atomic and molecular  gas reservoirs are not significantly depleted by the merging
process, do galaxy mergers lead to the cessation of star formation
in their post-merger remnant? If the answer is `yes' (e.g. Wiegel
et al. 2017), then the mechanism may not be gas consumption or
expulsion as previously proposed.  Instead, there may be a
mechanism that prevents the cold gas reservoir (which we have
shown be particularly abundant in post-mergers) from turning
into stars.  Indeed, the 0.2 dex excess in
atomic (this paper) and molecular (Sargent et al., in prep) gas
fractions, even when matching in SFR, does suggest that conversion
of these two phases is proceeding at a less efficient rate in post-mergers
(see also Davis et al. 2015).

The internal dynamics of galactic gas may offer a plausible
mechanism for decreasing the star formation efficiency of merging
galaxies.  Although our understanding of gas instabilities is over half
a century old (Safronov 1960; Toomre 1964),
there is a growing appreciation that kinematics
may play a crucial role in governing the state
of galactic gas and the potential for quenching in a galaxy
(e.g.  Obreschkow et al. 2016; Wong et al. 2016;
Pontzen et al. 2017; Falgarone et al. 2017).
The ISM of galaxy mergers may plausibly be expected to be highly turbulent,
due to a combination of gravitational torques, shocks and outflows from
star formation and AGN (e.g. Veilleux et al. 2013; Sell et al. 2014;
Rich, Kewley \& Dopita 2015; Mortazavi \& Lotz 2018). 
In turn, it is expected that increased turbulence will make the ISM
stable against gravitational collapse, even if there is abundant
gas present in the galaxy (e.g. Alatalo et al. 2015; Smercina et al.
2018).  van de Voort et al. (2018)
have recently claimed that such dynamical effects may explain the
low star formation efficiencies in a sample of early-type post-merger
galaxies, despite significant cold gas reservoirs.  For the quenching
to be long-lived, the injection of turbulence would need to be persistent
(or recurrent); stochastic AGN activity in the `maintenance mode'
is one possible mechanism to achieve this (e.g. Pontzen et al. 2017).

Alternatively, the gas remaining after the merger's starburst might simply be
too diffuse to efficiently form stars. Maps of star formation, atomic and
molecular gas in nearby galaxies have shown that it is the latter that
correlates with star formation, rather than the former (e.g. Bigiel et al. 2008;
Leroy et al. 2008, 2013).  Moreover, it is important to recognize that the
molecular gas traced by CO 1-0 observations (the transition used in most
of the molecular gas studies referenced herein) can be relatively diffuse
and not necessarily in compact giant molecular clouds (e.g.
Pety et al. 2013; Caldu-Primo et al. 2015).  Stars form out of the
\textit{dense} molecular gas that is better traced with transitions such
as HCN (e.g. Wu et al. 2005; Lada et al. 2012).
Even within the dense gas the star formation
efficiency and dense gas fraction seem to vary on local conditions
(Usero et al. 2015;
Bigiel et al. 2016).  Taken taken, these studies of gas in local
galaxies show that an abundance of gas (even an abundance of molecular
gas) is not a sufficient condition to form stars, with a complex dependence
on local conditions.

\subsection{A revised view of the galaxy merger process}

In Fig. \ref{evo_fig} we suggest a revision of the Hopkins et al. (2008) schematic
picture of the galaxy merger process, pieced together from, and illustrated by, a decade
of work on mergers selected from the SDSS and complementary multi-wavelength data. 

\begin{enumerate}

\item  Galaxies that experience a close encounter trigger tidal disturbances that
morphologically alter both the stellar and gas components (e.g. Ellison
et al. 2010; Patton et al. 2016).  The interaction leads to
diluted central metallicities, enhanced SFRs (e.g. Ellison et al. 2008; 
Patton et al. 2011) and elevated AGN fractions (Ellison et al. 2011;
Satyapal et al. 2014; Khabiboulline et al. 2014). 

\item  As the galaxies
separate after their first pericentric passage, changes in the
metallicities, asymmetries, SFR and AGN fractions can persist to wide separations
(e.g. Scudder et al. 2012; Patton et al. 2013, 2016).  

\item Close to coalescence, dual AGN become more frequent (Ellison et al. 2011, 2013).
The increased nuclear obscuration by gas and dust result in a higher fraction
of AGN that are selected in the mid-IR (Satyapal et al. 2014), a realization
that has resulted in a significant increase in the number of dual AGN with
separations $<$ 10 kpc (Ellison et al. 2017; Satyapal et al. 2017).

\item After coalescence, recent post-merger galaxies exhibit SFRs and AGN
fraction enhancements that exceed the pre-merger phase (Ellison et al. 2013).
Atomic and molecular gas fractions are elevated for their stellar mass and SFR
(this work; Sargent et al. in prep).

\item  We speculate that the abundant atomic and molecular gas in the 
post-merger remnant is neither
consumed by star formation, nor expelled by winds.  Complementary observations
of post-merger and post-starburst galaxies suggest that both atomic and molecular
gas reservoirs are retained, but
rendered (at least temporarily) infertile either due to increased shocks and turbulence 
(e.g. Buyle et al. 2006;  Zwaan et al. 2013; French et al. 2015; Rowlands et al. 2015; Alatalo et al. 2016;
Suess et al. 2017; van de Voort et al. 2018; Smercina et al. 2018) or to
its diffuse distribution/lower dense gas fraction.

\end{enumerate}

Although many aspects of this picture remain the same as that proposed by Hopkins et
al. (2008), (e.g. triggered star formation and nuclear accretion),
we have found no evidence for the `blowout' phase.

\section{Conclusions}\label{conclusions_sec}

We have compiled a sample of 107 visually classified post-merger
galaxies within a fixed right ascension and declination footprint
(10 $< \alpha <$ 17 hours, 0 $< \delta <$ 37 degrees) with
$z<0.04$ and log M$_{\star} > 9.0$ M$_{\odot}$.  47 of these post-mergers have
existing measurements of \mhi\ in the literature. Arecibo observations 
of 51 of the remaining 60 post-mergers have been completed 
with an observing strategy identical to the xGASS sample of $\sim$ 1200
representative galaxies from from the SDSS with log M$_{\star} > 9.0$
(Catinella et al. 2018).  The identical observing strategy,
data reduction pipeline and survey depth between xGASS and
the post-mergers, facilitate a robust comparison between the
two samples.  The work presented here represents a factor of
$\sim$ 3 increase in sample size and a factor of $\sim$ 5
in depth compared to our previous assessment of post-merger
atomic gas fractions in Ellison et al. (2015).

\medskip\

Our primary results are as follows.

\begin{itemize}
  
\item  \textbf{Post-mergers exhibit an elevated \HI\ detection fraction
  compared with xGASS galaxies at the same stellar mass.}
  We define an effective detection
threshold $f_{gas}$ = 2 percent for log M$_{\star} >$ 10.5 M$_{\odot}$
and log \mhi = 8.8  M$_{\odot}$ below that mass.
The fraction of post-merger galaxies above this detection threshold exceeds
that of the xGASS sample at fixed stellar mass.  For log M$_{\star} <$
10.5 M$_{\odot}$, the mass regime where most of our post-mergers are located,
the \HI\ detection fraction is enhanced by $\sim$ 50 per cent relative
to the xGASS sample (Fig \ref{det_frac}).

\medskip

\item \textbf{The median \HI\ gas fraction is larger in the post-merger sample
  than in xGASS at fixed stellar mass.}  When only considering the \HI\
  detections, the median \fgas\ in the post-merger sample is larger than
  in xGASS by $\sim$ 0.2 dex. However, due to the different detection fractions
  in the two samples, this is an under-estimate of the true difference
  in median gas fractions.  Accounting for the \HI\ non-detections, the median
  \fgas\ in the post-merger sample is 0.3 -- 0.6 dex larger than in
  xGASS (Fig. \ref{fgas_med}).

\medskip

\item  \textbf{The median atomic gas fraction enhancement in post-mergers is
  a factor of 3 higher than control galaxies in xGASS.}  In order to compute the atomic gas fraction
  offset on a galaxy-by-galaxy basis, we construct a mass matched control 
  sample of xGASS galaxies for each post-merger.  Accounting for non-detections
  the median \dfgas\ = 0.51 dex (Fig. \ref{dfgas}).  However, if SFR is
additionally included in the matching parameters, the median atomic gas fraction
enhancement is reduced to  \dfgas\ = 0.2 dex (Fig. \ref{dfgas_v5}).
  
\medskip

\item \textbf{There is no correlation between a galaxy's \HI\ gas fraction enhancement
and its SFR enhancement (Fig. \ref{dsfr}) or asymmetry.}  
There is also no difference in the distribution
of \dfgas\ for galaxies hosting an AGN.

\medskip

\item \textbf{We find no evidence for an environmental origin for the enhanced atomic
gas fractions in post-mergers.}  \dfgas\ values are re-computed after requiring that
control galaxies have the same halo classification (satellite, group central,
isolated central) as the post-merger galaxy.  Both group and isolated centrals
in the post-merger sample have similar positive distributions of \dfgas\
(Fig. \ref{dfgas_halo}).

\end{itemize}

We conclude that mergers represent an effective mechanism 
for elevating the \HI\ gas fraction at a given stellar mass.  However, 
the lack of merger signatures in the general xGASS galaxy sample 
with high \fgas\ indicates that mergers are not the only way for a 
galaxy to cultivate an \HI\ rich ISM. We propose 
that the elevated \fgas\ in post-mergers could be
a combination of the arithmetic combination of the progenitor galaxies,
and the net cooling of the ionized/hot gas component of the ISM/halo.

Simulations by Moreno et al. (2018) find that atomic
gas reservoirs remain relatively constant throughout the merger
process. Combined with the results presented here,
and other observational findings that post-starburst galaxies
retain significant gas reservoirs (French et al. 2015; Rowlands
et al. 2015; Suess et al. 2017), we conclude that
merger-induced star formation is unlikely to lead to quenching
via gas exhaustion or expulsion.  However, the enhanced
ISM turbulence that is likely to follow a galaxy merger
could decrease the star formation efficiency (e.g. van de Voort
et al. 2018), or lower the dense gas fraction and ultimately facilitate the transition to
a passive remnant.  

\section*{Acknowledgements}

SLE gratefully acknowledges support from a CAASTRO visiting fellowship to
UWA, during which this work was initiated, and from
an NSERC Discovery Grant.
BC is the recipient of an Australian Research Council Future Fellowship (FT120100660). 
Parts of this research were conducted by the Australian Research Council Centre of 
Excellence for All Sky Astrophysics in 3 Dimensions (ASTRO 3D), through project number 
CE170100013. Our thanks to Jorge Moreno and Mark Sargent for allowing us to cite
results prior to publication.

This research has made use of the NASA/IPAC Extragalactic Database
(NED) which is operated by the Jet Propulsion Laboratory, California
Institute of Technology, under contract with the National Aeronautics
and Space Administration.

The Arecibo Observatory is operated by SRI International under a
cooperative agreement with the National Science Foundation
(AST-1100968), and in alliance with Ana G. M{\'e}ndez-Universidad
Metropolitana, and the Universities Space Research Association.

Funding for the SDSS and SDSS-II has been provided by the Alfred P. Sloan Foundation, the Participating Institutions, the National Science Foundation, the U.S. Department of Energy, the National Aeronautics and Space Administration, the Japanese Monbukagakusho, the Max Planck Society, and the Higher Education Funding Council for England. The SDSS Web Site is http://www.sdss.org/.

The SDSS is managed by the Astrophysical Research Consortium for the Participating Institutions. The Participating Institutions are the American Museum of Natural History, Astrophysical Institute Potsdam, University of Basel, University of Cambridge, Case Western Reserve University, University of Chicago, Drexel University, Fermilab, the Institute for Advanced Study, the Japan Participation Group, Johns Hopkins University, the Joint Institute for Nuclear Astrophysics, the Kavli Institute for Particle Astrophysics and Cosmology, the Korean Scientist Group, the Chinese Academy of Sciences (LAMOST), Los Alamos National Laboratory, the Max-Planck-Institute for Astronomy (MPIA), the Max-Planck-Institute for Astrophysics (MPA), New Mexico State University, Ohio State University, University of Pittsburgh, University of Portsmouth, Princeton University, the United States Naval Observatory, and the University of Washington.

\appendix

\section{Arecibo spectra}

Fig. \ref{detect_fig} shows the first 12 \HI\ detections from Table \ref{obs_table}.
In Figs. \ref{app_det_fig} and \ref{app_nd_fig} we show the remaining detections and non-detections
respectively.

\clearpage
\begin{figure*}
\begin{center}
\includegraphics[width=16cm]{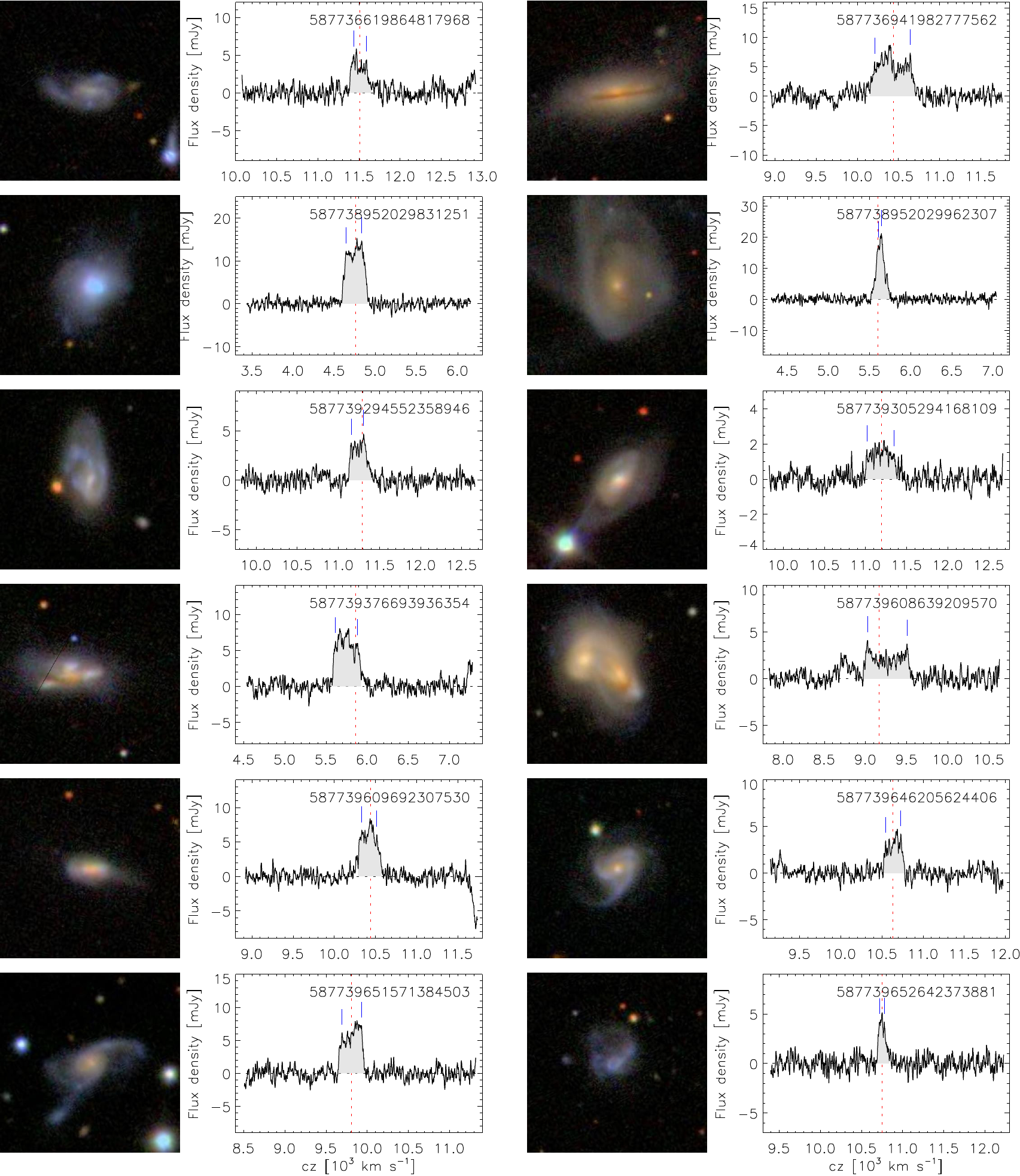}
\caption{SDSS postage stamp images (1 arcmin square) and \HI -line profiles of the galaxies detected in this work, for the 
remaining galaxies in Table \ref{obs_table}, ordered by SDSS objID (as indicated in the panels). 
The \HI\ spectra are calibrated, smoothed and baseline-subtracted. A dotted line and two dashes indicate the 
heliocentric velocity corresponding to the SDSS redshift and the two peaks used for width measurement,
respectively.}
\label{app_det_fig}
\end{center}
\end{figure*}

\clearpage
\setcounter{figure}{0}
\begin{figure*}
\begin{center}
\includegraphics[width=16cm]{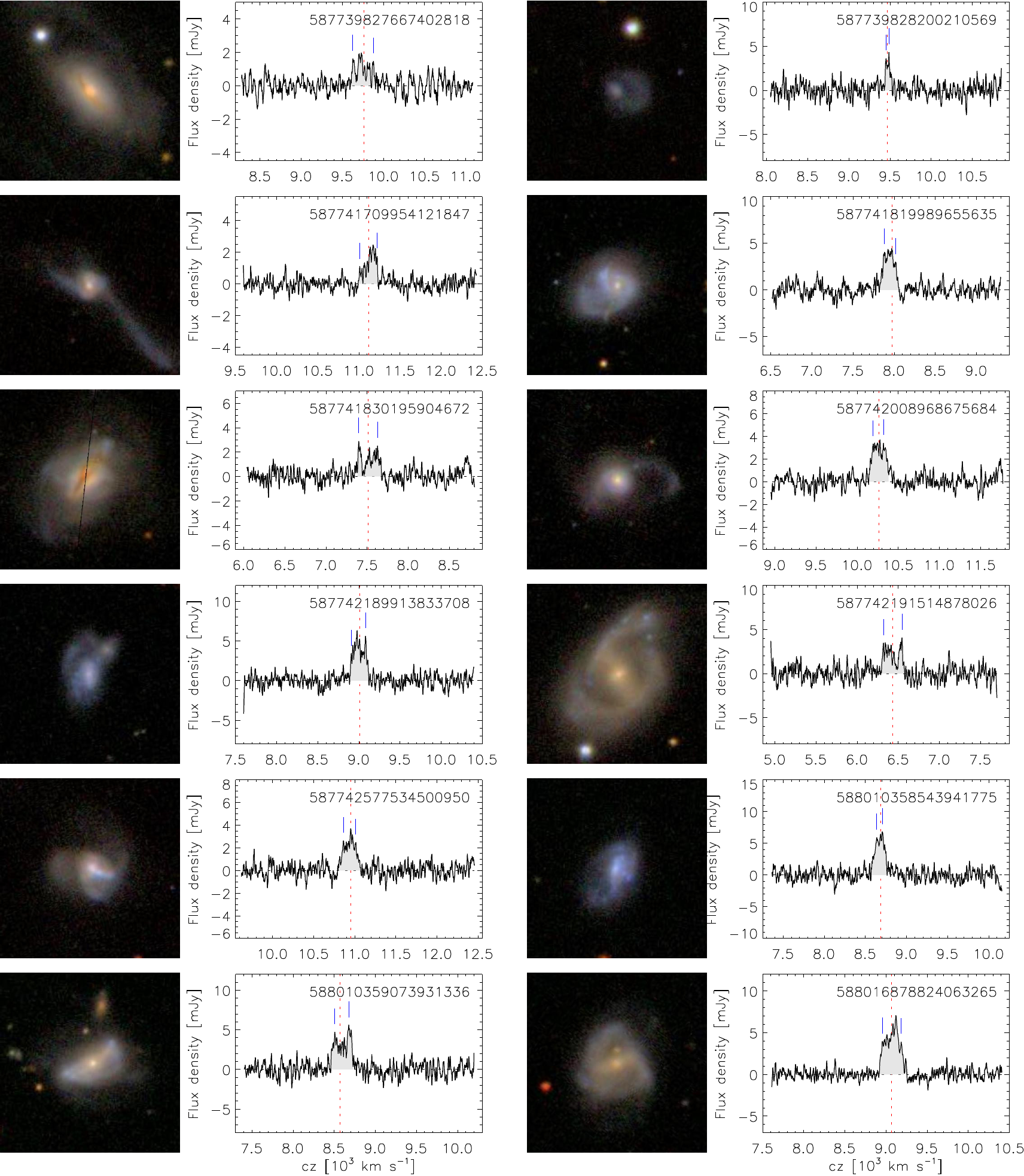}
\caption{\it continued}
\end{center}
\end{figure*}

\clearpage
\setcounter{figure}{0}
\begin{figure*}
\begin{center}
\includegraphics[width=16cm]{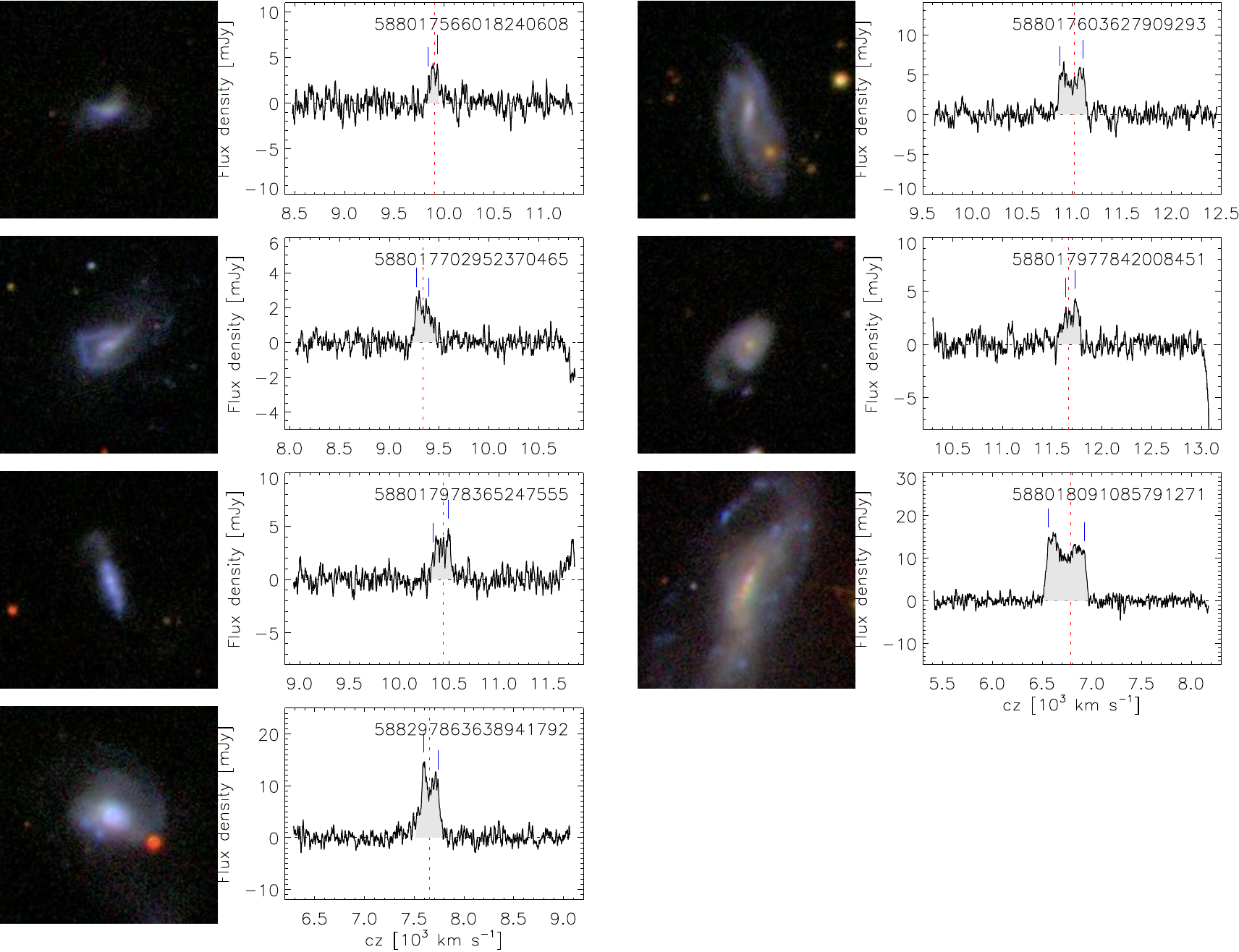}
\caption{\it continued}
\end{center}
\end{figure*}

\begin{figure*}
\includegraphics[width=16cm]{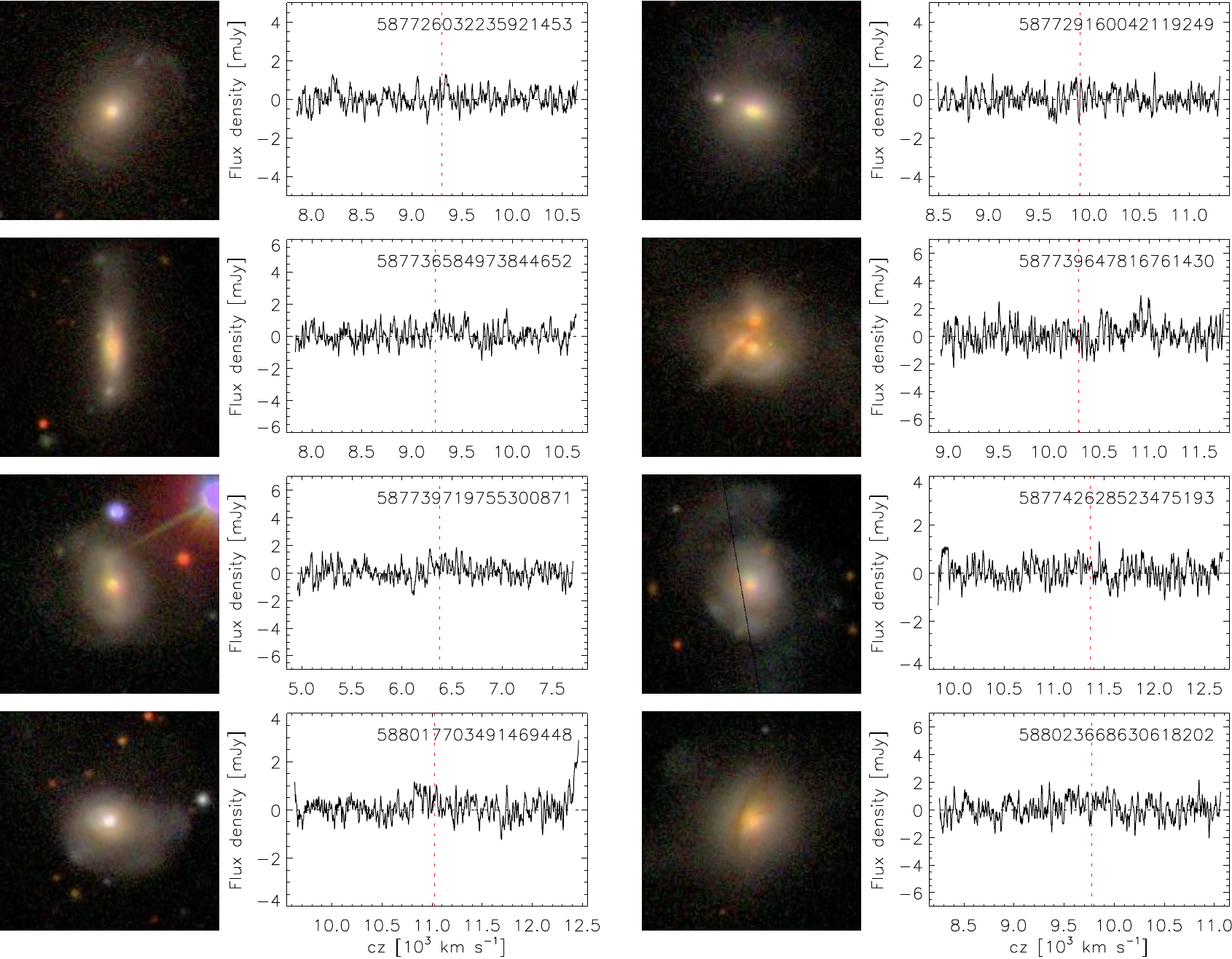}
\caption{Same as Fig. \ref{app_det_fig} for non-detections.}
\label{app_nd_fig}
\end{figure*}


\begin{thebibliography}{}
\small
\itemindent -0.48cm

\bibitem[Aalto et al. 2001]{aa01}
        Aalto, S., Huttemeister, S., Polatidis, A. G., 2001, A\&A, 372, L29

\bibitem[Alatalo et al. (2015)]{ala15} 
        Alatalo, K, et al., 2015, ApJ, 812, 117

\bibitem[Alatalo et al. (2016)]{ala16} 
        Alatalo, K, et al., 2016, ApJ, 827, 106

\bibitem[Bekki (1998)]{bek98}
        Bekki, K., 1998, ApJ, 502, L133

\bibitem[Berg et al. (2018)]{cosagn}
        Berg, T. A. M.,  Ellison, S. L., Tumlinson, J., Oppenheimer, B.,
	Horton, R., Bordoloi, R., Schaye, J., 2018, MNRAS, in press

\bibitem[Bigiel et al. (2008)]{big08}
        Bigiel, F., Leroy, A., Walter, F., Brinks, E., de Blok, W. J. G., 
	Madore, B., Thornley, M. D.,  2008, AJ, 136, 2846

\bibitem[Bigiel et al. (2016)]{big16}
        Bigiel, F., et al. 2016, ApJ, 822, L26

\bibitem[Blecha et al. (2018)]{ble18}
        Blecha, L., Snyder, G. F., Satyapal, S., Ellison, S. L.,
	2018, MNRAS, submitted	

\bibitem[Braine \& Combes (1993)]{bc93}
         Braine, J., \& Combes, F., 1993, A\&A, 269, 7

\bibitem[Brinchmann et al. 2004]{bri04} 
        Brinchmann, J., Charlot, S., White, S. D. M., Tremonti, C., 
	Kauffmann, G., Heckman, T., Brinkmann, J.,2004, MNRAS, 351, 1151  

\bibitem[Bustamante et al. (2018)]{bus18}
         Bustamante, S., Sparre, M., Springel, V., Grand, R. J. J.,
	 2018, MNRAS, submitted

\bibitem[Buyle et al. (2006)]{buy06}
        Buyle, P., Michielsen, D., De Rijcke, S., Pisano, D. J.,
	Dejonghe, H., Freeman, K.,  2006, ApJ, 649, 163

\bibitem[Caldu-Primo et al. 2015]{cp15}
         Caldu-Primo, A., Schruba, A., Walter, F., Leroy, A., Bolatto,
         A. D., Vogel, S., 2015, AJ, 149, 76
        
\bibitem[Capelo et al. (2015)]{cap15}
        Capelo, P. R., Volonteri, M., Dotti, M., Bellovary, J. M.,
	Mayer, L., Governato, F.,  2015, MNRAS, 447, 2123

\bibitem[Casasola et al. (2004)]{cas04}
         Casasola, V., Bettoni, D., Galletta, G., 2004, A\&A, 422, 941	

\bibitem[Catinella et al. (2010)]{cat10}
        Catinella, B., et al,
	2010, MNRAS, 403, 683

\bibitem[Catinella et al. (2012)]{cat12}
        Catinella, B., et al, 2012, A\&A, 544, 65

\bibitem[Catinella et al. (2013)]{cat13}
        Catinella, B., et al, 2013, MNRAS, 436, 34

\bibitem[Catinella et al. (2018)]{cat18}
        Catinella, B., et al, 2018, MNRAS, 476, 875

\bibitem[Catinella \& Cortese (2015)]{cc15}
         Catinella, B., \& Cortese, L., 2015, MNRAS, 446, 3526

\bibitem[Chung et al. (2009)]{viva}
        Chung, A., van Gorkom, J. H., Kenney, J. D. P., Crowl, H., 
	Vollmer, B., 2009, AJ, 138, 1741

\bibitem[Combes et al. (1994)]{com94}
        Combes, F., Prugniel, P., Rampazzo, R., Sulentic, J. W.,
	1994, A\&A, 281, 725

\bibitem[Cortese et al. (2011)]{cor11}
        Cortese, L., Catinella, B., Boissier, S., Boselli, A., Heinis, S.,
	2011, MNRAS, 415, 1797

\bibitem[Cox et al. (2008)]{cox08}
        Cox, T. J., Jonsson, P., Somerville, R. S., Primack, J. R.,
	Dekel, A., 2008, MNRAS,  384, 386

\bibitem[Darg et al. (2010a)]{darg10a} 
        Darg, D. W., et al., 2010, MNRAS, 401, 1552

\bibitem[Davis et al. (2015)]{dav15}
         Davis, T. A., et al.,  2015, MNRAS, 449, 3503

\bibitem[Di Matteo et al. (2007)]{dim07}
        Di Matteo, P., Combes, F., Melchior, A.-L., Semelin, B.,
	2007, A\&A, 468, 61

\bibitem[Di Matteo et al. (2005)]{dim05}
        Di Matteo, T., Springel, V., Hernquist, L., 2005, Nature, 433, 604

\bibitem[Ellison et al. (2008)]{sle08a}
         Ellison, S. L., Patton, D. R., Simard, L., McConnachie, A. W.,
	 2008 AJ, 135, 1877

\bibitem[Ellison et al. (2010)]{sle10} 
         Ellison, S. L., Patton, D. R., Simard, L., McConnachie, A. W.,
	 Baldry, I. K., Mendel, J. T.,
	 2010, MNRAS 407, 1514. 

\bibitem[Ellison et al. (2011)]{agn}
         Ellison, S. L., Patton, D. R.,  Mendel, J. T., Scudder, 
	 J. M., 2011, MNRAS, 418, 2043

\bibitem[Ellison et al. (2013)]{post} 
         Ellison, S. L., Mendel, J. T.,  Patton, D. R., Scudder, J. M.,
	2013, MNRAS, 453, 3627

\bibitem[Ellison et al., (2015)]{HI}
        Ellison, S. L., Fertig, D., Rosenberg, J. L., Nair, P., 
	Simard, L., Torrey, P., Patton, D. R., 2015, MNRAS, 448, 221

\bibitem[Ellison et al., (2017)]{dual}
        Ellison, S. L., Secrest, N. J., Mendel, J. T., Satyapal, S., 
	Simard, L., 2017, MNRAS, 470, L49

\bibitem[Ellison et al., (2018)]{manga}
        Ellison, S. L., Sanchez, S. F., Ibarra-Medel, H., Antonio, B.,
	Mendel, J. T., Barrera-Ballesteros, J., 2018, MNRAS, 474, 2039

\bibitem[Fabello et al. (2011)]{fab11}
         Fabello, S., Kauffmann, G., Catinella, B., Giovanelli, R., 
	 Haynes, M. P., Heckman, T. M., Schiminovich, D.,
	 2011, MNRAS, 416, 1739

\bibitem[Falgarone et al. (2017)]{fal}
        Falgarone, E., et al., 2017, Nature, 548, 430

\bibitem[Fernandez et al. (2010)]{fer10}
        Fernandez, X., van Gorkom, J. H., Schweizer, F., Barnes, J. E.,
	2010, AJ, 140, 1965

\bibitem[French et al. (2015)]{kdf15}
        French, K. D., Yang, Y., Zabludoff, A., Narayanan, D.,
	Shirley, Y., Walter, F., Smith, J.-D., Tremonti, C. A.,
	2015, ApJ, 801, 1

\bibitem[Georgakakis et al. (2000)]{geo00}
        Georgakakis, A., Forbes, D. A., Norris, Ray P.,
	2000, MNRAS, 318, 124

\bibitem[Gereb et al. (2018)]{ger18}
        Gereb, K., Janowiecki, S., Catinella, B., Cortese, L., Kilborn, V.,
	2018, MNRAS, 476, 896

\bibitem[Giovanelli et al. (2005)]{gio05a}
        Giovanelli, R., et al., 2005, AJ, 130, 2598
     
\bibitem[Goto (2005)]{goto05} 
        Goto, T., 2005, MNRAS, 357, 937

\bibitem[Goulding et al. (2018)]{gou18}
        Goulding, A. D., et al., 2018, PASJ, 70, 37

\bibitem[Hani et al. (2018)]{hani18}
        Hani, M., Sparre, M., Ellison, S. L., Torrey, P. Vogelsberger, M.,
	2018, MNRAS, in press

\bibitem[Haynes et al. (2011)]{hay11}
        Haynes, M. P., et al., 2011, AJ, 142, 170

\bibitem[Hess \& Wilcots (2013)]{hw13}
        Hess, K. M., Wilcots, E. M., 2013, AJ, 146, 124


\bibitem[Hibbard \& van Gorkom (1996)]{hvg96}
        Hibbard, J. E., van Gorkom, J. H., 1996, AJ, 111, 655	

\bibitem[Hibbard \& Yun (1999)]{hy99}
        Hibbard, J. E., \& Yun, M. S., 1999, AJ, 118, 162

\bibitem[Hopkins et al. (2008)]{hop08}
       Hopkins, P. F., Hernquist, L., Cox, T. J., Keres, D.,
       2008, ApJS, 175, 356

\bibitem[Hopkins et al. (2018)]{fire2}
        Hopkins, P. F., et al.,  2018, MNRAS, submitted

\bibitem[Huchtmeier et al. (2008)]{huc08}
        Huchtmeier, W. K., Petrosian, A., Gopal-Krishna, McLean, B., 
	Kunth, D., 2008, A\&A, 492, 367

\bibitem[Janowiecki et al. (2017)]{jano17}
        Janowiecki, S., Catinella, B., Cortese, L., Saintonge, A., 
	Brown, T., Wang, J.,  2017, MNRAS, 466, 4795

\bibitem[Jaskot et al. (2015)]{jas15}
        Jaskot, A. E., Oey, M. S., Salzer, J. J., Van Sistine, A., 
	Bell, E. F., Haynes, M. P.,  2015, ApJ, 808, 66

\bibitem[Ji, Peirani \& Yi (2014)]{jpy14}
        Ji, I., Peirani, S., \& Yi, S. K., 2014, A\&A, 566, 97

\bibitem[Kauffmann et al. (2003)]{kau03} 
	Kauffmann, G., et al., 2003, MNRAS, 346, 1055

\bibitem[Khabiboulline et al. (2014)]{emil14}
        Khabiboulline, E. T., Steinhardt, C. L., Silverman, J. D.,
	Ellison, S. L., Mendel, J. T., Patton, D. R.,
	 2014, ApJ, 795, 62

\bibitem[Kilborn et al. (2009)]{kil09}
         Kilborn, V. A., Forbes, D. A., Barnes, D. G., Koribalski, B. S., 
	 Brough, S., Kern, K., 2009, MNRAS, 400, 1962

\bibitem[Koribalski \& Dickey (2014)]{kd04}
        Koribalski, B., Dickey, J. M.,	
	2004, MNRAS, 348, 1255

\bibitem[Larson et al. 2016]{lar16}
        Larson, K. L., et al.,  2016, ApJ, 825, 128

\bibitem[Leroy et al. (2008)]{ler08}
        Leroy, A. K., Walter, F., Brinks, E., Bigiel, F., de Blok, W. J. G.,
	Madore, B., Thornley, M. D., 2008, AJ, 136, 2782

\bibitem[Leroy et al. (2013)]{ler13}
        Leroy, A. K., et al., 2013, AJ, 146, 19

\bibitem[Lada, Lombardi \& Alves (2010)]{lla10}
         Lada, C. J., Forbrich, J., Lombardi, M.,  \& Alves, J. F.,
         2012, ApJ, 745, 190

\bibitem[Lotz et al. (2010)]{lotz10a} 
        Lotz, J. M., Jonsson, P., Cox, T. J., Primack, J. R.,
	2010a, MNRAS, 404, 590

\bibitem[Lotz et al. (2010)]{lotz10b} 
        Lotz, J. M., Jonsson, P., Cox, T. J., Primack, J. R.,
	2010b, MNRAS, 404, 575

\bibitem[Lutz (2018)]{lutz18}
        Lutz, K. A., et al., 2018, MNRAS, 476, 3744

\bibitem[Manthey et al. (2008)]{man08a}
       Manthey, E., Aalto, S., Huttemeister, S., Oosterloo, T. A.,
       2008a, A\&A, 484, 693

\bibitem[Manthey et al. (2008)]{man08b}
       Manthey, E., Huttemeister, S.,  Aalto, S., Horellou, C., Bjerkeli, P.,
       2008b, A\&A, 490, 975

\bibitem[Meusinger et al. (2017)]{meu}
        Meusinger, H., Brunecke, J., Schalldach, P., in der Au, A.,
	2017, A\&A, 597, 134

\bibitem[Michiyama et al. (2016)]{mich16}
         Michiyama, T., et al.,  2016, PASJ, 68, 96

\bibitem[Mihos \& Hernquist (1994)]{mh94}
        Mihos, C., \& Hernquist, L., 1994, ApJ, 425, L13

\bibitem[Mihos \& Hernquist (1996)]{mh96}
        Mihos, C., \& Hernquist, L., 1996, ApJ, 464, 641

\bibitem[Montuori et al. (2010)]{mon10}
         Montuori, M., Di Matteo, P., Lehnert, M. D., Combes, F., 
	 Semelin, B.,  2010, A\&A, 518, 56

\bibitem[Mortazavi \& Lotz (2018)]{mort19}
         Mortazavi, S. A., \& Lotz, J. M., 2018, MNRAS, submitted

\bibitem[Mortazavi et al. (2016)]{mort16}
         Mortazavi, S. A., Lotz, J. M., Barnes, J. E., Snyder, G. F.,
         2016, MNRAS, 455, 3058
         
\bibitem[Moster et al. (2011)]{mos11}
        Moster, B. P., Maccio, A. V., Somerville, R. S., Naab, T., Cox, T. J.
	 2011, MNRAS, 415, 3750

\bibitem[Moreno et al. (2015)]{mor15}
        Moreno, J., Torrey, P., Ellison, S. L., Patton, D. R., 
	Bluck, A. F. L., Bansal, G., Hernquist, L., 2015, MNRAS, 448, 1107

\bibitem[Moreno et al. (2018)]{mor18}
        Moreno, J., Torry, P., Ellison, S. L., et al., 2018, MNRAS, submitted       

\bibitem[Nair \& Abraham (2010a)]{na10a}
        Nair, P. B.,  \& Abraham, R. G., 2010, ApJS, 186, 427

\bibitem[Obreschkow et al. (2016)]{ob16}
        Obreschkow, D., Glazebrook, K., Kilborn, V., Lutz, K.,
	2016, ApJ, 824, L26O

\bibitem[Patton et al. (2011)]{dave11}
         Patton, D. R., Ellison, S. L.,  Simard, L., McConnachie, A. W.,
	 Mendel, J. T.,
	 2011, MNRAS, 412, 591 

\bibitem[Patton et al. (2013)]{dave13}
        Patton, D. R., Torrey, P., Ellison, S. L., Mendel, J. T., 
	Scudder, J. M., 2013, MNRAS, 433, L59

\bibitem[Patton et al. (2016)]{dave16}
         Patton, D. R., Qamar, F. D., Ellison, S. L.,  Bluck, A. F. L.,
	 Simard, L., Mendel, J. T., Moreno, J., Torrey, P.,
	 2016, MNRAS, 461, 2589

\bibitem[Pawlik et al. (2018)]{paw18}
        Pawlik, M. M., et al., 2018, MNRAS, in press          
         
\bibitem[Perez et al. (2011)]{per11} 
        Perez, J., Michel-Dansac, L., Tissera, P. B.,
	2011, MNRAS,  417, 580

\bibitem[Pety et al. (2013)]{pet13}
        Pety, J., et al., 2013, ApJ, 779, 43
        
\bibitem[Pontzen et al. (2017)]{pon17}
        Pontzen, A., et al., 2017, MNRAS, 465, 547

\bibitem[Pop et al. (2018)]{pop}
         Pop, A.-R., Pillepich, A., Amorisco, N. C., Hernquist, L.,
         2018, MNRAS, submitted
        
\bibitem[Rafieferantsoa et al. (2015)]{rafi15}
        Rafieferantsoa, M., Dave, R., Angles-Alcazar, D., Katz, N., 
	Kollmeier, J. A., Oppenheimer, B. D., 2015, MNRAS, 453, 3980

\bibitem[Renaud et al. (2014)]{ren14}
        Renaud, F., Bournaud, F., Kraljic, K., Duc, P.-A.,
	2014, MNRAS, 442, L33

\bibitem[Rich et al. (2015)]{rich15}
        Rich, J. A.,  Kewley, L. J., Dopita, M. A.,
	2015, ApJS, 221, 28

\bibitem[Rowlands et al. (2015)]{row15}
        Rowlands, K., Wild, V., Nesvadba, N., Sibthorpe, B., 
	Mortier, A., Lehnert, M., da Cunha, E., 2015, MNRAS, 448, 258

\bibitem[Rupke, Kewley \& Barnes (2010)]{rkb10} 
        Rupke, D. S. N., Kewley, L. J., Barnes, J. E.,
	 2010, ApJ, 710, L156

\bibitem[Safronov (1960)]{saf60}
        Safronov, V. S., 1960, AnAp, 23, 979

\bibitem[Saintonge et al. (2012)]{sain12}
         Saintonge, A., et al., 2012, ApJ, 758, 73

\bibitem[Saintonge (2016)]{sain16}
        Saintonge, A., et al., 2016, MNRAS, 462, 1749

\bibitem[Salim et al. (2007)]{sal07}
        Salim, S., et al, 2007, ApJS, 173, 267

\bibitem[Sancisi et al. (2008)]{san08}
         Sancisi, R., Fraternali, F., Oosterloo, T., van der Hulst, T.,
	 2008, A\&ARv, 15, 189

\bibitem[Satyapal et al. (2014)]{sat14} 
        Satyapal, S., Ellison, S. L., McAlpine, W., Hickox, R. C., 
	Patton, D. R., Mendel, J. T., 2014, MNRAS, 441, 1297

\bibitem[Satyapal et al. (2017)]{sat17} 
        Satyapal, S.,  et al., 2017, ApJ, 848, 126

\bibitem[Schawinski et al. (2015)]{sch15}
         Schawinski, K., Koss, M., Berney, S., Sartori, L. F.,
	  2015, MNRAS, 451, 2517

\bibitem[Scudder et al. (2012b)]{[pairs}
         Scudder, J. M., Ellison, S. L., Torrey, P., Patton, D. R.,
	 Mendel, J. T., 2012, MNRAS, 426, 549

\bibitem[Sell et al. (2014)]{sell14}
        Sell, P. H. et al., 2014, MNRAS, 441, 3417

\bibitem[Simard et al. (2011)]{sim11}
        Simard, L., Mendel, J. T., Patton, D. R., Ellison S. L., 
	McConnachie, A. W., 2011, ApJS, 196, 11

\bibitem[Smercina et al. (2018)]{sme18}
        Smercina, A., et al., 2018, ApJ, 855, 51

\bibitem[Solanes et al. (2001)]{sol01}
         Solanes, J. M., Manrique, A., Garcia-Gomez, C., Gonzalez-Casado, G., 
	 Giovanelli, R., Haynes, M. P.,   2001, ApJ, 548, 97

\bibitem[Solomon \& Sage (1988)]{ss88}
        Solomon, P. M., \& Sage, L. J.,  1988, ApJ, 334, 613

\bibitem[Springob et al. (2005)]{sp05}
        Springob, C. M., Haynes, M. P., Giovanelli, R., Kent, B. R.,

\bibitem[Stierwalt et al., 2015]{tnt15}
        Stierwalt, S., Besla, G., Patton, D., Johnson, K., Kallivayalil, N., 
	Putman, M., Privon, G., Ross, G., 2015, ApJ, 805, 2

\bibitem[Suess et al. (2017)]{sue17}
        Suess, K. A., Bezanson, R., Spilker, J. S., Kriek, M.,
	Greene, J. E., Feldmann, R., Hunt, Q., Narayanan, D.,
	2017, ApJ, 846, L14

\bibitem[Tacconi et al. (1999)]{tac99}
        Tacconi, L. J., Genzel, R., Tecza, M., Gallimore, J. F.,
	Downes, D., Scoville, N. Z., 1999, ApJ, 524, 732

\bibitem[Teyssier et al. 2010)]{tey10}
        Teyssier, R., Chapon, D., Bournaud, F.,  2010, ApJ, 720, L149

\bibitem[Tonnesen \& Cen (2012)]{tc12}
        Tonnesen, S., \& Cen, R., 2012, MNRAS, 425, 2313

\bibitem[Toomre (1964)]{too64}
         Toomre, A., 1964, ApJ, 139, 1217

\bibitem[Torrey et al. (2012)]{torr2012}
         Torrey, P., Cox, T. J., Kewley, L. J., Hernquist, L.,
	  2012, ApJ, 746, 108

\bibitem[Ueda et al. (2014)]{ueda14}
         Ueda, J., et al. 2014, ApJS, 214, 1

\bibitem[Usero et al., (2015)]{us15}
        Usero, A., et al., 2015, AJ, 150, 115

\bibitem[van de Voort et al. (2018)]{vdv18}
        van de Voort, F., et al., 2018, MNRAS, 476, 122

\bibitem[Veilleux et al. (2013)]{v13}
        Veilleux, S., et al., 2013, ApJ, 776, 27

\bibitem[Violino et al. 2018]{vio}
        Violino, G., Ellison, S. L., Sargent, M., Coppin, K. E. K., 
	Scudder, J. M., Mendel, J. T.,  Saintonge, A.,
	MNRAS, 2018, 476, 2591

\bibitem[Wang et al. (2004)]{wang04}
        Wang, J., Zhang, Q., Wang, Z., Ho, P. T. P., Fazio, G. G., Wu, Y.,
	 2004, ApJ, 616, L67

\bibitem[Weigel et al. (2017)]{wei17}
        Weigel, A. K., et al.,  2017, ApJ, 845, 145

\bibitem[Weston et al. (2017)]{wes17}
        Weston, M. E., McIntosh, D. H., Brodwin, M., Mann, J., Cooper, A., 
	McConnell, A., Nielsen, J. L.,  2017, MNRAS, 464, 3882

\bibitem[Woods \& Geller (2007)]{wg07}
        Woods, D. F., Geller, M. J., 
	2007, AJ, 134, 527

\bibitem[Wong et al. (2016)]{wong16}
         Wong, O. I., Meurer, G. R., Zheng, Z., Heckman, T. M.,
	 Thilker, D. A., Zwaan, M. A., 2016, MNRAS, 460, 1106

\bibitem[Wu et al. 2005]{wu05}
         Wu, J., Evans, N. J., II, Gao, Y., Solomon, P. M., Shirley, Y. L.,
         Vanden Bout, P. A., 2005, ApJ, 635, L173
         
\bibitem[Yamashita et al. (2017)]{yam17}
        Yamashita, T., et al., 2017, ApJ, 844, 96

\bibitem[Yang et al. (2007)]{yang07}
        Yang, X., Mo, H. J., van den Bosch, F. C., Pasquali, A., 
	Li, C., Barden, M., 2007, ApJ, 671, 153

\bibitem[Young et al. (2014)]{you14}
        Young, L. M., et al., 2014, MNRAS, 444, 3408

\bibitem[Yun \& Hibbard (2001)]{yh01}
        Yun, M. S., \& Hibbard, J. E., 2001, ApJ, 550, 104

\bibitem[Zabludoff et al. (1996)]{zab96}
        Zabludoff, A. I., Zaritsky, D., Lin, H., Tucker, D., 
	Hashimoto, Y., Shectman, S. A., Oemler, A., Kirshner, R. P.,
	1996, ApJ, 466, 104

\bibitem[Zuo et al. (2018)]{zuo18}
        Zuo, P., Xu, C. K., Yun, M. S., Lisenfeld, U., Li, D., Cao, C.,
	2018, ApJ, in press

\bibitem[Zwaan et al. (2013)]{zwa13}
        Zwaan, M. A., Kuntschner, H., Pracy, M. B., Couch, W. J.,
	2013, MNRAS, 432, 492

\end{thebibliography}
\end{document}